\documentclass[authoryear, review]{elsarticle}

\usepackage{natbib}
\setcitestyle{authoryear,open={[},close={]}}

\usepackage[utf8]{inputenc}
\usepackage{float}
\usepackage[nolist]{acronym}

\usepackage{amsmath,amssymb,amsfonts,enumerate}

\usepackage{placeins}
\let\Oldsection\section
\renewcommand{\section}{\FloatBarrier\Oldsection}
\let\Oldsubsection\subsection
\renewcommand{\subsection}{\FloatBarrier\Oldsubsection}
\let\Oldsubsubsection\subsubsection
\renewcommand{\subsubsection}{\FloatBarrier\Oldsubsubsection}

\usepackage{array}
\newcolumntype{P}[1]{>{\centering\arraybackslash}p{#1}}
\newcolumntype{L}[1]{>{\raggedright\arraybackslash}p{#1}}

\usepackage{multirow}
\usepackage{pbox}
\usepackage{rotating}
\usepackage{makecell}
\usepackage{graphicx}
\usepackage{booktabs}
\usepackage{comment}
\usepackage{adjustbox}
\usepackage{listings}
\usepackage{tikz}
\usetikzlibrary{positioning}
\usepackage{caption}

\usepackage{booktabs}

\usepackage{xcolor}

\colorlet{punct}{red!60!black}
\definecolor{background}{HTML}{EEEEEE}
\definecolor{delim}{RGB}{20,105,176}
\colorlet{numb}{magenta!60!black}

\definecolor{eclipseStrings}{RGB}{42,0.0,255}
\definecolor{eclipseKeywords}{RGB}{127,0,85}
\definecolor{eclipseKeywords}{RGB}{0,0,0}

\lstdefinelanguage{json}{
    basicstyle=\linespread{0.8}\notsotiny,
    numberstyle=\scriptsize,
    commentstyle=\color{eclipseStrings},
    stringstyle=\color{eclipseKeywords},
    numbers=left,
    stepnumber=1,
    numbersep=8pt,
    showstringspaces=false,
    breaklines=true,
    frame=lines,
    backgroundcolor=\color{background},
    string=[s]{"}{"},
    comment=[l]{:\ "},
    morecomment=[l]{:"},
    literate=
        *{0}{{{\color{numb}0}}}{1}
         {1}{{{\color{numb}1}}}{1}
         {2}{{{\color{numb}2}}}{1}
         {3}{{{\color{numb}3}}}{1}
         {4}{{{\color{numb}4}}}{1}
         {5}{{{\color{numb}5}}}{1}
         {6}{{{\color{numb}6}}}{1}
         {7}{{{\color{numb}7}}}{1}
         {8}{{{\color{numb}8}}}{1}
         {9}{{{\color{numb}9}}}{1}
}

\makeatletter
\newcommand\notsotiny{\@setfontsize\notsotiny{6}{8}}
\makeatother

\begin{acronym}
\acro{dlt}[DLT]{Distributed Ledger Technologies}
\acro{dl}[DL]{Distributed Ledger}
\acro{bc}[BC]{Blockchain}
\acroplural{dl}[DLs]{Distribted Ledgers}
\acro{dlps}[DLPS]{Distributed Ledger Performance Scan}
\acro{evm}[EVM]{Ethereum Virtual Machine}
\acro{pow}[PoW]{Proof-of-Work}
\acro{pos}[PoS]{Proof-of-Stake}
\acro{poa}[PoA]{Proof-of-Authority}
\acro{cft}[CFT]{Crash Fault Tolerance}
\acro{bft}[BFT]{Byzantine Fault Tolerance}
\acro{pbft}[PBFT]{Practical \ac{bft}}
\acro{rbft}[RBFT]{Redundant \ac{bft}}
\acro{ibft}[IBFT]{Istanbul \ac{bft}}
\acro{poet}[PoET]{Proof of Elapsed Time}
\acro{raft}[RAFT]{Reliable, Replicated, Redundant, And Fault-Tolerant}
\acro{tee}[TEE]{Trusted Execution Environment}
\acro{aws}[AWS]{Amazon Web Services}
\acro{p2p}[P2P]{Peer-to-Peer}
\acro{sc}[SC]{Smart Contract}
\acroplural{sc}[SCs]{Smart Contracts}
\acro{couchdb}[CDB]{CouchDB}
\acro{leveldb}[LDB]{LevelDB}
\acro{dlt}[DLT]{Distributed Ledger Technologies}
\acro{dl}[DL]{Distributed Ledger}
\acroplural{dl}[DLs]{Distribted Ledgers}
\acro{dlps}[DLPS]{Distributed Ledger Performance Scan}
\acro{evm}[EVM]{Ethereum Virtual Machine}
\acro{pow}[PoW]{Proof-of-Work}
\acro{pos}[PoS]{Proof-of-Stake}
\acro{poa}[PoA]{Proof-of-Authority}
\acro{cft}[CFT]{Crash Fault Tolerance}
\acro{bft}[BFT]{Byzantine Fault Tolerance}
\acro{pbft}[PBFT]{Practical \ac{bft}}
\acro{rbft}[RBFT]{Redundant \ac{bft}}
\acro{ibft}[IBFT]{Istanbul \ac{bft}}
\acro{poet}[PoET]{Proof of Elapsed Time}
\acro{raft}[RAFT]{Reliable, Replicated, Redundant, And Fault-Tolerant}
\acro{tee}[TEE]{Trusted Execution Environment}
\acro{aws}[AWS]{Amazon Web Services}
\acro{p2p}[P2P]{Peer-to-Peer}
\acro{sc}[SC]{Smart Contract}
\acroplural{sc}[SCs]{Smart Contracts}
\acro{couchdb}[CDB]{CouchDB}
\acro{leveldb}[LDB]{LevelDB}
\acro{hlf}[Fabric]{Hyperledger Fabric}
\end{acronym}

\PassOptionsToPackage{hyphens}{url}\usepackage[hidelinks]{hyperref}

\usepackage{lineno}
\modulolinenumbers[5]

\journal{Computers \& Industrial Engineering}

\newif\ifblinded
\blindedfalse

\newif\ifcoverpage
\coverpagetrue

\newif\ifdraft
\drafttrue
\ifdraft
\definecolor{ocolor}{rgb}{1,0,0.4}
\newcommand{\alnote}[1]{ {\textcolor{red} { ***André: #1 }}}
\newcommand{\jsnote}[1]{ {\textcolor{blue} { ***Johannes: #1 }}}
\newcommand{\tgnote}[1]{ {\textcolor{green} {***Tobi: #1 }}}
\else
\newcommand{\alnote}[1]{}
\newcommand{\prnote}[1]{}
\newcommand{\jsnote}[1]{}
\newcommand{\tgnote}[1]{}
\newcommand{\note}[1]{}
\fi

\renewcommand{\arraystretch}{1.25} 

\usepackage{pdfrender}

\usepackage{pifont}
\newcommand{\cmark}{\ding{51}}%
\newcommand{\xmark}{\ding{55}}%
\AtBeginDocument{%
  \providecommand\BibTeX{{%
    \normalfont B\kern-0.5em{\scshape i\kern-0.25em b}\kern-0.8em\TeX}}}

\begin{document}

\title{An In-Depth Investigation of the Performance Characteristics of Hyperledger Fabric}


\ifcoverpage

\author{\texorpdfstring{\mbox{Tobias Guggenberger\,{$^{\mathrm{a},\mathrm{b},\dagger}$}}}}
\author{\texorpdfstring{\mbox{Johannes Sedlmeir\,{$^{\mathrm{a},\mathrm{b},\dagger,\ast}$}}}}
\author{\texorpdfstring{\mbox{Gilbert Fridgen\,{$^{\mathrm{b},\mathrm{c}}$}}}}
\author{\texorpdfstring{\mbox{Andre Luckow\,{$^{\mathrm{d}}$}}}}

\address{
    $^\mathrm{a}$
    Branch Business \& Information Systems Engineering of Fraunhofer FIT, Bayreuth, Germany\\
    $^\mathrm{b}$
    FIM Research Center, University of Bayreuth, Bayreuth, Germany\\
    $^\mathrm{c}$
    SnT - Interdisciplinary Center for Security, Reliability and Trust, University of Luxembourg, Luxembourg City, Luxembourg\\
    $^\mathrm{d}$
    BMW Group, Munich, Germany \\~\\
    $^\dagger$ 
    Both first authors contributed equally to the article \\
    $^\ast$
    Corresponding author: \href{mailto:johannes.sedlmeir@fim-rc.de}{johannes.sedlmeir@fim-rc.de}}

\else
\author{}
\address{}
\fi

\begin{frontmatter}

\begin{abstract}

Private permissioned blockchains are deployed in ever greater numbers to facilitate cross-organizational processes in various industries, particularly in supply chain management. One popular example of this trend is Hyperledger Fabric. Compared to public permissionless blockchains, it promises improved performance and provides certain features that address key requirements of enterprises. However, also permissioned blockchains are still not as scalable as centralized systems, and due to the scarcity of theoretical results and empirical data, their real-world performance cannot be predicted with the necessary precision. We intend to address this issue by conducting an in-depth performance analysis of Hyperledger Fabric. The paper presents a detailed compilation of various performance characteristics using an enhanced version of the Distributed Ledger Performance Scan (DLPS). Researchers and practitioners alike can use the various performance properties identified and discussed as guidelines to better configure and implement their Hyperledger Fabric network. Likewise, they are encouraged to use the DLPS framework to conduct their measurements.

\end{abstract}

\end{frontmatter}

\ifcoverpage
\pagenumbering{gobble} 
\section*{Acknowledgment}
We thank Philipp Ross, Emil Djerekarov, Jannik Lockl, and Daniel Miehle for their support in creating the first version of the DLPS. Our gratitude further extends to Colin Glass, Marius Poke, and Orestis Papageorgiou for their valuable comments.
This research is supported by the Bavarian Ministry of Economic Affairs, Regional Development and Energy. We gratefully acknowledge their funding of the project ``Fraunhofer Blockchain Center (20-3066-2-6-14)''. This research was also funded in part by the Luxembourg National Research Fund (FNR) and PayPal, grant reference “P17/IS/13342933/PayPal-FNR/Chair in DFS/Gilbert Fridgen” (PEARL) and the FiReSpARX Project (C20/IS/14783405/FiReSpARX/Fridgen).
\clearpage
\else
\fi

\pagenumbering{gobble} 
\noindent\textbf{Highlights}
\begin{itemize}
    \item Performance is a widely acknowledged challenge for industrial blockchains
    \item Numerous parameters impact Fabric's maximum throughput; latency is consistently low.
    \item Hardware, database, network size, and privacy can have a high impact on throughput.
    \item Fabric offers good performance even in large intercontinental supply chain networks.

\end{itemize}

\clearpage
\textbf{Graphical Abstract}
\begin{figure}[H]
    \centering
    \includegraphics[width=0.9\textwidth]{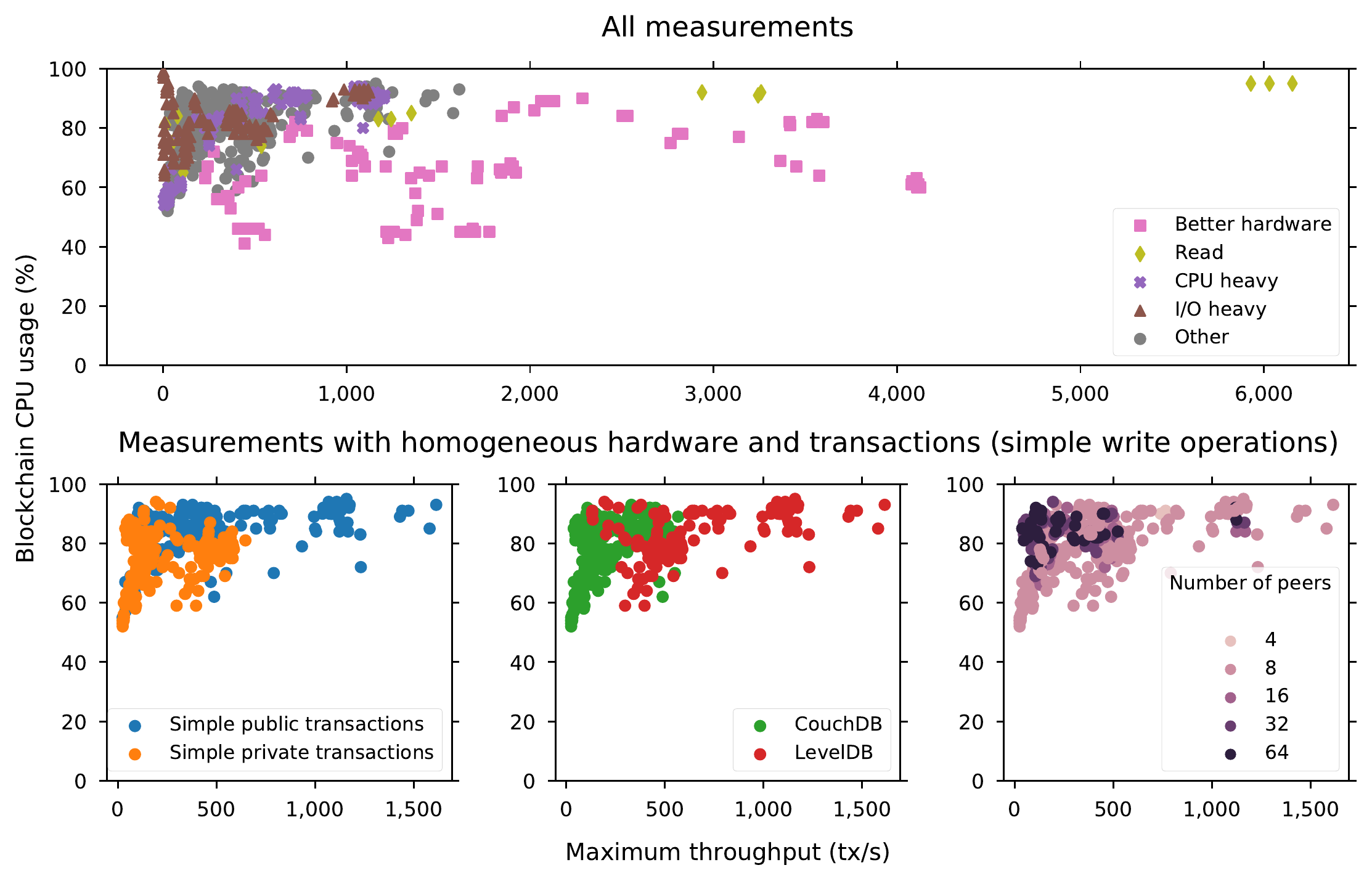}
    \caption*{This chart illustrates influential experimental parameters on the throughput of Hyperledger Fabric networks. In this paper, we illustrate that multiple parameters have a significant impact on performance metrics. Our results were collected by setting up more than~1,500~Hyperledger Fabric networks and operating more than 200~million transactions in experiments that ran for more than~2,000 hours. The purpose of these efforts is to validate and extend previous research by evaluating more than 15~network- and transaction-related parameters, including choices of hardware and database, transaction payloads and privacy configurations, different network sizes of 5 to 128 nodes, and geographic distribution. We also analyze the impact of node crashes.}
    \label{fig:abstract}
\end{figure}

\clearpage


\clearpage
\section{Introduction}
\label{sec:introduction}

Bitcoin was the first blockchain system, developed primarily for the decentralization of financial systems. It created a new virtual currency that allows transactions without the involvement of distinct intermediaries like banks~\citep{Nakamoto:2008:Bitcoin}. Based on this general concept,~\citet{Buterin:2014:Ethereum} extended the scope of blockchain technology to a broader field of application. At that time, blockchain technology was only able to provide users with limited programming logic, so Ethereum improved its versatility by introducing a programming language and an associated runtime environment (``Ethereum virtual machine'') to execute smart contracts. These were first conceptualized by~\citet{szabo1997formalizing}, and even in their early form, they facilitated the execution of highly customizable program code in a \ac{p2p} environment without relying on a distinct intermediary. The advancement of blockchain technology has fostered the development of decentralized applications in the business and indeed in the public sector, while they have gone far beyond the initial use cases in the financial sector~\citep{labazova2019hype,casino2019systematic} for cross-organizational workflows~\citep{fridgen2018cross}. This progress extends to applications in the public sector~\citep{rieger2019building}, the pharmaceutical sector~\citep{Mattke:2019:BlockchainPharma}, and the automotive sector~\citep{miehle2019partchain}. Especially within supply chain management, many researchers agree that blockchain provides a viable infrastructure that facilitates a more efficient way of sharing information, improved data quality, and traceable product provenance~\citep{agrawal2021blockchain,azzi2019power,guggenberger2020improving,lim2021literature,longo2019blockchain,reddy2021developing,sunny2020supply,jensen2019tradelens}.

In their comprehensive literature review of current developments and potential applications of blockchain in supply chain management, \citet{chang2020blockchain} conclude that blockchain has the potential to disrupt supply chain operations and provide not only distributed governance and process automation but also improved performance across the board. Yet despite the considerable benefits that distributed ledgers can offer enterprises, such as consolidating audit and production data in an unimpeachable distributed database, public blockchains are still subject to numerous limitations. Examples include their generally rather high transaction fees, their lack of finality, and their inability to safeguard transaction confidentiality \citep{kannengiesser2020,sedlmeir2022serverless}. Many permissionless blockchains are also based on Proof of Work, which is why they are extremely high in energy consumption \citep{sedlmeir2020energy}; an inconvenient truth that is difficult to reconcile with corporate sustainability goals. Fortunately, we were able to note a broad awareness of the challenges concerning throughput and latency. In conducting a systematic study of the literature on blockchain-based supply chain applications, we found that 83 of 128 publications acknowledged performance challenges. Meanwhile, privacy issues were noted by 71, security issues by 60, regulatory issues by 47, and challenges concerning costs by 41. To name but one example, \citet{perboli2018blockchain} analyzed a supply chain use case and mentioned multiple times that performance evaluation is a crucial step in the design and implementation of blockchain-based supply chain solutions.

In attempts to resolve these issues and answer the increasing demand for enterprise-level blockchain applications expressed in various industries, developers have introduced new frameworks and modified blockchain architectures. These are intended to compensate for the shortcomings of public permissionless blockchains and adapt them to the needs of enterprises \citep{kannengiesser2020}. To achieve these goals, frameworks were developed in such a way as to implement private permissioned blockchains that restrict participation in the blockchain and consensus to a consortium \citep{beck2018governance}. While other blockchains like Hyperledger Sawtooth have also been examined with regard to their potential to support supply chain applications \citep{perboli2020blockchain}, \ac{hlf} has arguably become the preeminent technical support structure in this domain. Indeed, \citet{guggenberger2020improving}, \citet{lim2021literature}, and \citet{reddy2021developing} have all discussed the use of \ac{hlf} for large-scale cross-enterprise applications, mainly because the framework provides high security and performance as well as flexible tools for managing access, safeguarding privacy, and implementing business logic \citep{androulaki2018hyperledger,kannengiesser2020}.

At present, several major projects based on \ac{hlf} are transitioning from tests and minimum viable products with limited scope to production-ready systems, as a result of which there is a growing number of participating parties and operations in these projects \citep{ibm2020blockchain,miehle2019partchain}. When looking at these projects, however, it becomes apparent that the requirements concerning private or public transactions, the varying complexities of smart contracts, and the need to support and adapt network topologies all differ significantly \citep{kannengiesser2020}. It is a valuable asset, therefore, that \ac{hlf} offers various configurations which allow one to adapt it to a wide range of different use case requirements \citep{10.1145/3366370}. This is especially significant in projects like TradeLens~\citep{jensen2019tradelens} since its purpose is to provide the infrastructure for worldwide supply chain monitoring, which means that it places extensive requirements on the performance of blockchain systems. The choice of various architectural parameters, such as network size, hardware configuration, internet connection speed, and complexity of operations (i.e., smart contracts methods), is known to have a large impact on the performance of distributed systems in general and in particular on that of blockchains ~\citep[see, e.g., ][]{Baliga:2018:Quorum,Thakkar:2018:Fabric}. It is inevitable, then, that trade-offs between security, network size, privacy, and performance must be considered when designing a system with high performance and reliability requirements \citep{kannengiesser2020}.

In Section~\ref{sec:relatedWork} of this paper, our literature review identifies two significant gaps in the current body of knowledge on the general performance of permissioned blockchains and the specific performance of \ac{hlf}. The first of these gaps is the result of the fact that, to date, studies have focused on particular variables without allowing for a holistic view, mainly because these studies have conducted their measurements with non-standardized tools. Furthermore, many have allowed for ambiguity in the definition of their key metrics and indeed in the attainment of their results. Therefore, multiple observations are neither replicable nor generalizable \citep{sedlmeir2021dlps}, and until they are, there can be no holistic view of the performance of \ac{hlf}. As for the second research gap, this is the result of the fact that \ac{hlf} has been developing at a considerable pace, frequently offering new configuration options and features that impact its performance at a rate too fast and extensive to have been covered by the limited literature to date. For example, since private data collections can provide a certain level of access control in a cross-enterprise system, they are essential for many  enterprise-level applications~\citep{10.1145/3316481,10.1145/3366370,sedlmeir2022transparency}. The private data transaction process, however, is far more complex than the conventional transaction process because it introduces additional gossip routines. These protocol changes make it difficult to predict the performance of private data transactions compared to that of conventional transactions, and yet, to the best of our knowledge, there has been no rigorous academic study of the many ways in which the use of private data collections can impact performance.

In this paper, we intend to close these research gaps by studying a wide variety of \ac{hlf}’s performance characteristics. Our in-depth analysis thereof comprises the perspectives of both researchers and architects of large-scale enterprise and public sector projects. Our measurements significantly extend the range of performance characteristics studied to date, including additional scenarios that are highly relevant to the real-world use of blockchain technology, e.g., in supply chain applications in the industrial and public sectors. As~\citet{kannengiesser2020} have pointed out, the right balance of these factors is essential if one’s deployment of blockchain is to facilitate the most effective creation of value. Our research objective, therefore, is to develop a list of relevant variables, measure their specific impact on different \ac{hlf} implementations, and demonstrate the potential of \ac{hlf} in various scenarios. In doing so, we hope to advance the understanding of enterprise blockchains and what they represent; a highly complex fault-tolerant distributed system applied in real-world settings, as required by enterprises. To be more specific, we put the spotlight on the capabilities of \ac{hlf}, one of the blockchain frameworks most frequently used by industry consortia. While \ac{hlf} has been discussed by~\citet{lim2021literature} and~\citet{reddy2021developing} as an infrastructure for supply chain systems supported by blockchain, we provide further insights into how private permissioned blockchains can support large-scale and indeed global supply chains. Our findings include that \ac{hlf} scales exceedingly well with CPU-heavy transactions but struggles with transaction payloads larger than 100 kB. Furthermore, whereas \ac{hlf} is very suitable for intercontinental networks, private transactions, in particular, suffer from commensurate high latency. Since our overarching research aim is to close many of the gaps in the knowledge on the performance characteristics of blockchain, this paper also provides an extension of the \ac{dlps}~\citep{sedlmeir2021dlps}, a blockchain benchmarking framework. The \ac{dlps} offers not only clear definitions of key performance metrics but also an end-to-end description of their setup and measurement, which ensures full transparency and repeatability. We supply our extension of the \ac{dlps} in the open-source repository~\citep{DLPS} along with the results of our experiments so that researchers can repeat our measurements or use them to easily examine new configurations.

The remainder of this paper is structured as follows: Section~\ref{sec:background} gives an overview of the key concepts on which \ac{hlf} and its architecture are predicated. Section~\ref{sec:relatedWork} provides a thorough review of the literature on benchmarking \ac{hlf} and identifies the main gaps to be closed. Section~\ref{sec:methods} describes the measurement process involved in the \ac{dlps} in detail. Section~\ref{sec:results} then presents the main findings of this study by demonstrating our benchmarking results under careful consideration of the wide range of variables employed in the benchmark tests. In Section~\ref{sec:discussion}, we discuss our findings, outline their implications for real-world applications, and provide design guidelines. Finally, in Section~\ref{sec:conclusion}, we identify opportunities for future research.

\section{Hyperledger Fabric: Technical Background}
\label{sec:background}

\subsection{System Architecture}

\begin{figure}[!b]
    \centering
    \includegraphics[width=0.8\textwidth, trim=0cm 2.5cm 0cm 3.5cm, page=1, clip]{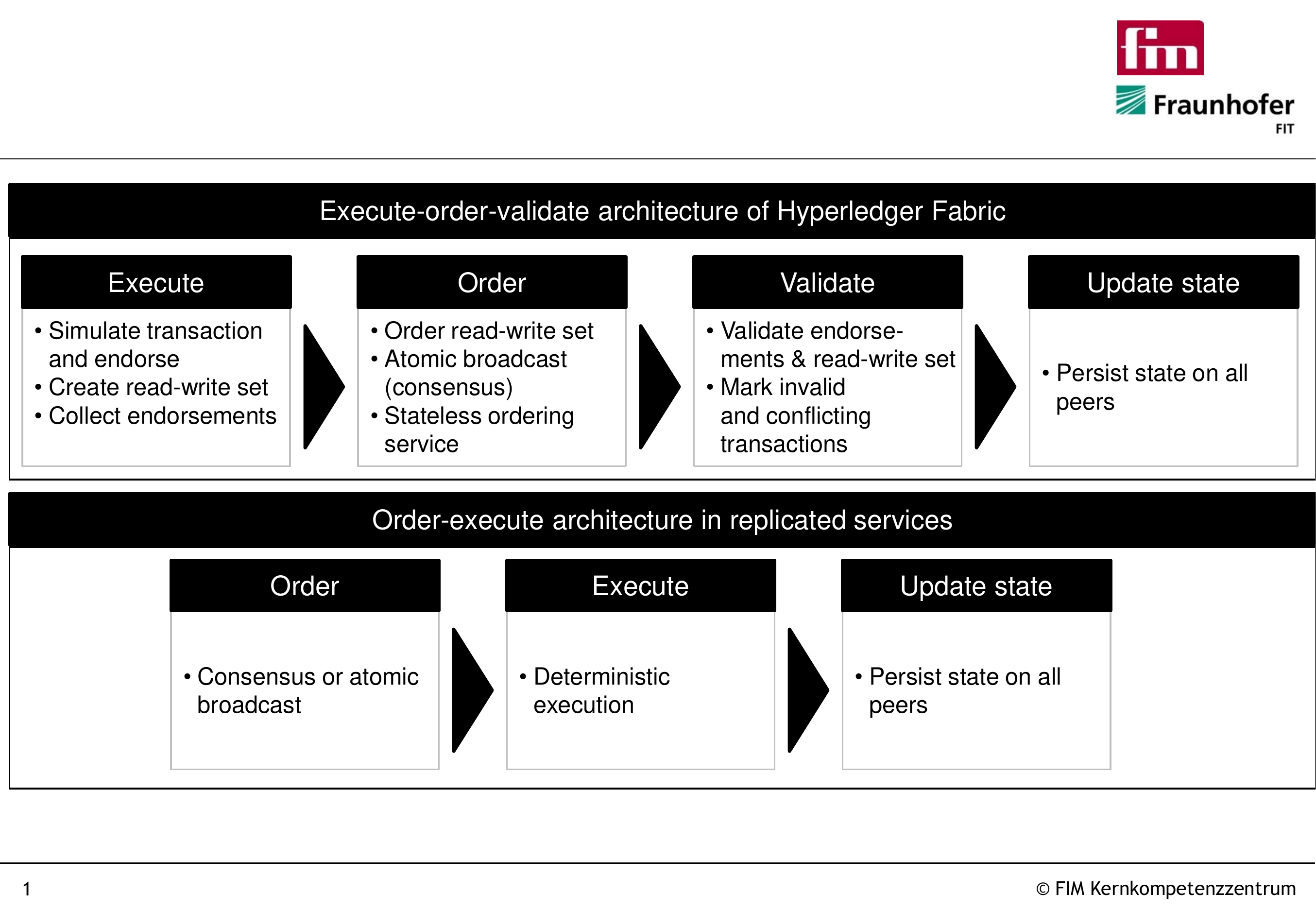}
    \caption{The execute-order-validate paradigm in \ac{hlf} compared to the order-execute architecture common to most blockchains.}
    \label{fig:hl_transaction_order}
\end{figure}

Since the days of version 1.0, \ac{hlf} has facilitated a paradigm that fundamentally differs from most blockchains in that it offers improved performance, flexibility, and privacy features~\citep{10.1145/3366370}. Instead of relying on an order-execute architecture, \ac{hlf} uses an execute-order-validate paradigm (see Figure~\ref{fig:hl_transaction_order}). Order-execute means two things: first, that the consensus mechanism is responsible for ordering and then broadcasting new transactions, and second, that all peers execute these transactions sequentially. In contrast, execute-order-validate implies that \ac{hlf} separates execution and validation from ordering~\citep{androulaki2018hyperledger}.

This altered replication process requires a new system architecture. A \ac{hlf} node can perform one of the following three roles~\citep{androulaki2018hyperledger}:

\begin{itemize}
    
    \item Clients are responsible for submitting a transaction proposal to their peers and broadcasting any transactions in the form of a bundled endorsement response to bring transactions into order~\citep{androulaki2018hyperledger}.
    
    \item Peers receive the transaction proposals from clients, simulate them, and send the signed result back to the clients. Eventually, they validate transactions. All peers maintain a ledger consisting of an append-only data structure (blockchain) of all previous transactions and a structure that represents the latest world state of the ledger. To store this ledger state, peers can use conventional databases. At present, \ac{hlf} 2.0 supports LevelDB and CouchDB. Due to the execute-order-validate paradigm, \ac{hlf} does not require all peers to execute all transaction proposals. This design feature sets \ac{hlf} apart from most other blockchains, be they public permissionless or private permissioned blockchains. By means of an endorsement policy, one can specify the subset of peers required for the transaction proposal execution, and this can be done individually for each smart contract method. These subsets of peers are also called endorsers or endorsing peers~\citep{androulaki2018hyperledger}.
    
    \item Ordering Service Nodes, also known as orderers, together form the ordering service. This ordering service is responsible for creating the total order of all transactions. There are different ways of implementing the ordering service, ranging from a (now deprecated) solo orderer to distributed protocols, such as RAFT~\citep{Ongaro:2014:RAFT} and Kafka~\citep{kreps2011kafka}. While these protocols already address different levels of fault tolerance~\citep{androulaki2018hyperledger}, developers are working on future ordering services that should also account for byzantine faults~\citep{Lamport:1982:Byzantine}.
    
\end{itemize}

Clients, peers, and orderers are further grouped into organizations (abbreviated as orgs). These typically represent companies or wider groups of participants. Based on their organizational affiliation, these entities have different rights, like the permission to join a blockchain channel that represents a private subnet of communication between two or more network participants including a corresponding ledger. A peer can either join one or multiple channels. Those who opt for a greater number of nodes in one organization will experience increased redundancy and, therefore, reduced efficiency and networking overhead. On the other hand, the distribution of the simulation workload to more servers approach facilitates parallelization and, thus, a higher throughput of endorsements of transactions~\citep{thakkar2021scaling}.

\subsection{Transaction Flow}
\label{subsec:general_flow}
\begin{figure}[!b]
    \centering
    \includegraphics[width=0.8\textwidth, trim=0cm 1.5cm 0cm 5cm, page=2, clip]{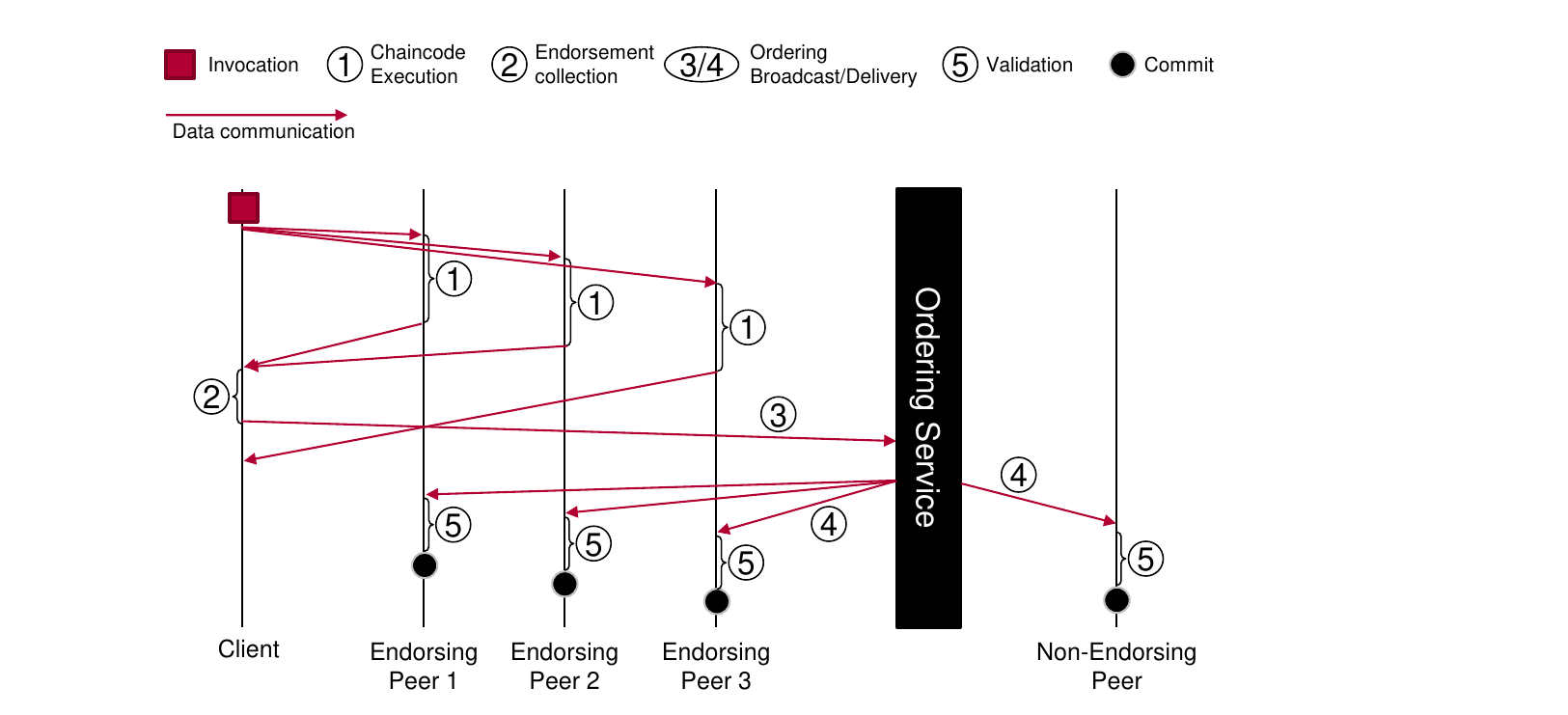}
    \caption{\ac{hlf} high-level transaction flow, adapted from~\citet{androulaki2018hyperledger}.}
    \label{fig:hl_transaction_flow}
\end{figure}

\ac{hlf}’s execute-order-validate paradigm separates the transaction flow into three parts: i) execution (sometimes also referred to as simulation) of a transaction, which involves checking its correctness by comparing the signed result of redundant execution on different peers. This is also called an endorsement. ii) ordering, which is done by means of a consensus protocol, regardless of the semantics of a transaction. iii) transaction validation, which ensures the endorsement policy and state consistency~\citep{androulaki2018hyperledger}. Figure~\ref{fig:hl_transaction_flow} provides an overview of the transaction flow.

\begin{enumerate}[(i)]

    \item\textbf{Execution Phase:} A client sends a cryptographically signed transaction proposal to one or more endorsing peers for execution (simulation). The peers do not yet update their ledger but only generate a read set and a write set~(1). The write set consists of all key updates resulting from the simulation, whereas the read set contains all keys that the peers read during the simulation. The endorsers then create a cryptographically signed endorsement, including the read and write sets, and send this back to the client in the form of a proposal response. The client collects endorsements until the requirements set by the endorsement policy are met~(2). This action further ensures that enough endorsers produce the same execution result and, in doing so, respond with the same read-write set~\citep{androulaki2018hyperledger}.

    \item\textbf{Ordering Phase:} Once the client has received enough consistent endorsements, the client bundles them all, creates a signed transaction, and sends it to the ordering service~(3). The ordering service uses consensus to establish a total order of all transactions. Furthermore, the ordering service batches the transactions in blocks and signs them~\citep{androulaki2018hyperledger}.

    \item\textbf{Validation Phase:} Blocks can either be delivered directly by the ordering service or indirectly by other peers who do so through a gossip protocol~(4). When a new block is delivered to a peer, it enters the validation phase~(5), which involves the following three sequential steps~\citep{androulaki2018hyperledger}:
    
    \begin{enumerate}[a.]
        
        \item The peer checks whether every transaction meets the endorsement requirements. If a transaction is invalid, the peer will mark it accordingly and ignore its effect.
        
        \item The peer enters the ledger update (``commit'') phase and appends the block to the local store ledger. For each transaction not marked as invalid, the peer writes all key-value pairs of the write set to the local state. Therefore, \ac{hlf} records invalid transactions even though they do not affect the state~\citep{androulaki2018hyperledger}.

    \end{enumerate}

\end{enumerate}

\subsection{Private Data Transaction Flow}
\label{subsec:private_flow}

Since version 1.2, \ac{hlf} also supports private data by means of private data collections. These represent privacy policies that determine which peers ought to process and store related data and which organizations should be able to access it~\citep{ma2019privacy}. Private data in particular benefits from the execute-order-validate paradigm and endorsement policies that do not require every peer to recompute every transaction in order to validate it. This feature makes it possible to conduct transactions where only a subset of the organizations participating on the \ac{hlf} blockchain store the actual data, while the remaining organizations only see the transaction hash, without relying on complex cryptographic techniques, such as Zero-Knowledge Proofs or Homomorphic Encryption~\citep{ibmGithub}.

\begin{figure}[!b]
    \centering
    \includegraphics[width=0.8\textwidth, trim=0cm 1.3cm 0cm 5cm, page=3, clip]{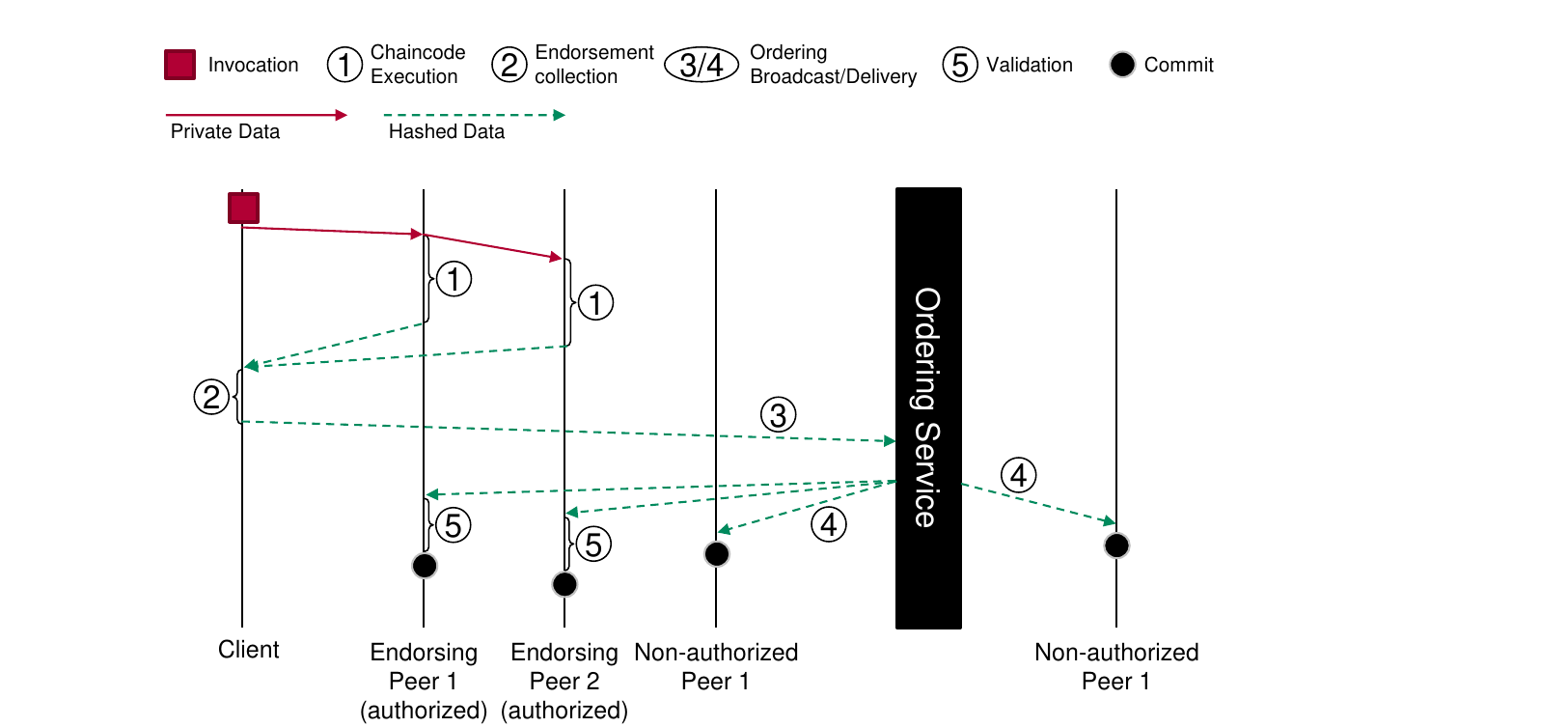}
    \caption{\ac{hlf} private data high-level transaction flow,
adapted from~\citet{androulaki2018hyperledger}.}
    \label{fig:hl_transaction_flow_private}
\end{figure}

Private data is mainly handled in accordance with the standard protocol discussed in Section~\ref{subsec:general_flow}, but at certain stages it departs from that protocol to ensure confidentiality in the three phases of execution, ordering, and validation (see Figure~\ref{fig:hl_transaction_flow_private}).

\begin{enumerate}[(i)]

    \item\textbf{Execution Phase:} 

    The client sends a proposal request, including the confidential data, to the designated endorser of the authorized organizations. Based on the collection policy that defines which organizations should be able to access this private data, the endorsing peers distribute it to other authorized peers via a gossip protocol. All peers in receipt of the private data store it in a transient data store. Similar to the general transaction flow, the endorsers generate a read-write set and send this to the client in the form of an endorsement~(1). These read-write sets do not contain any confidential data, however, but rather a hash of the private data keys and values. Once the client has received enough endorsements~(2), the client will send a transaction to the ordering service~(3) responsible for the total order of transactions~\citep{ma2019privacy}.

    \item\textbf{Ordering Phase:} The ordering phase follows a similar sequence as the general transaction flow. By consensus, the orderers include the transactions in a block and distribute them to all peers (4). Therefore, all peers receive the hashes of the private data, which facilitates subsequent validation~\citep{ma2019privacy}.

    \item\textbf{Validation Phase:} All peers will store the transaction in their ledger and update the read-write set with the associated hash values. Furthermore, in case a peer is authorized to access the private data related to the transaction, the peer will check the transient data store for the private data. If the peer has not received the private data in the execution phase, the peer will try to pull the private data from other authorized peers, whereupon the peer will use the hash values of the transaction to validate the private data and eventually save it to the private state database~\citep{ma2019privacy}. Generally, the peer makes use of one more table than the regular state database.
    
\end{enumerate}

While private data can be transferred confidentially between specific organizations, the required certificates are still used in plain text in order to verify permissions. Even though this is done without insight into the content of a transaction, this process entails severe confidentiality issues as it is apparent who is issuing new transactions. To safeguard against this confidentiality breach and hide the identity of the issuing client certificate, IBM introduced an implementation of the identity mixer protocol~\citep{bichsel2009cryptographic,camenisch2017practical}. However, not only is this feature highly experimental. To date, it is only supported in the Java-implementation of the \ac{hlf} client.

\section{Related Work}
\label{sec:relatedWork}

While the performance of blockchains is often considered crucial when working towards production-level systems~\citep{thakkar2021scaling}, at the time of writing, research in the field of systematic benchmarking is still scarce. To gain a full understanding of what research there is on the performance of \ac{hlf}, we conducted a structured literature review. To ensure the inclusion of all relevant publications, we first defined "Hyperledger AND Fabric" as a search string. We then used the string for queries in ACM Digital Library, AIS electronic Library, arXiv, IEEE Explore, and Web of Science. The initial set of search results totaled 1085 papers. After an initial screening of their titles and abstracts, we excluded 1007 publications for lack of relevance. Based on a subsequent full-text analysis, we removed every paper that performed benchmarking on a heavily modified version of \ac{hlf} as their findings are highly theoretical and not transferable to the publicly available versions of the system. After those eliminations, we were left with 19 articles that analyze the performance and scalability of \ac{hlf}. Table~\ref{table:literature_review_measurements} depicts this final set, the collective intelligence of which informed the following pages.

\begin{table}[!tb]
\centering
\resizebox{0.97\columnwidth}{!}{
\begin{tabular}{L{3cm}|p{19cm}}
\bf{Source} & \bf{Detailed content} \\\midrule

\citet{pongnumkul2017performance}. & This article presents a methodology for evaluating the performance of Ethereum and \ac{hlf}. The research team derives performance figures for execution time, latency, and throughput, while also considering various workloads.\\

\citet{androulaki2018hyperledger}. & This paper presents the execute-order-validate blockchain architecture of \ac{hlf} v1.1.0. The research team examines the throughput and latency under consideration of various parameters, such as block size, number of vCPUs, and number of peers.\\

\citet{baliga2018performance}. & This study makes use of Caliper to examine the performance of \ac{hlf} v1.0. The authors consider various influencing factors, including the number of nodes, endorsement policy, block size, and transaction size.\\

\citet{dinh2018untangling}. & The authors of this paper present the first systematic benchmarking framework for permissioned blockchains: Blockbench. It builds on the established YCSB and Smallbank frameworks to allow benchmarking of private Ethereum  (Geth, Parity), \ac{hlf}, and Quorum. With the use of this framework, the authors compare the performance of \ac{hlf} to Ethereum.\\

\citet{hao2018performance}. & This article presents a method of evaluating the performance of consensus algorithms in Ethereum and \ac{hlf}. The authors derive performance figures for latency and throughput, while also considering varying workloads.\\

\citet{nasir2018performance}. & The authors of this study compare the performances of \ac{hlf} v1.0 and v0.6. As well as analyzing execution time, latency, and throughput, they also vary the number of nodes to examine the scalability of the two implementations.\\

\citet{Thakkar:2018:Fabric}. & This study examines the impact of various factors on \ac{hlf} v1.1, such as block size, endorsement policy, channels, and state database choice. The authors identify performance bottlenecks and propose optimizations subsequently included in later versions of \ac{hlf}.\\

\citet{koushik2019performance}. & With the help of the Caliper benchmarking framework, the authors of this article investigate the performance of \ac{hlf} with regard to transaction throughput, latency, and send rate. They also analyze the impact of varying numbers of organizations.\\ 

\citet{kuzlu2019performance}. & Again with the help of Caliper, this research team examines the performance of \ac{hlf} with regard to throughput, response time, and simultaneous transactions.\\

\citet{nguyen2019impact}. & The authors of this study use a customized version of the Hyperledger Caliper benchmarking framework to examine the impact of sup-second network delays on the performance of \ac{hlf}. To create the \ac{hlf} network, they set up their test with two cloud instances, one in Germany and one in France.\\ 

\citet{dabbagh2020performance}. & The authors of this study use the Caliper benchmarking framework to compare the performance of \ac{hlf} with that of Ethereum. They also evaluate different versions of the \ac{hlf} SDK.\\

\citet{dreyer2020}. & For this research project, the authors analyze the performance of \ac{hlf} by creating various network configurations and measuring throughput, latency, and error rate, along with the overall scalability of the \ac{hlf} platform. The results are presented in the context of older versions of \ac{hlf}.\\ 

\citet{geneiatakis2020}. & The authors focus on the application of blockchain in the field of cross-border e-government services. However, they also take separate account of the performance of \ac{hlf}. Among other variables, they discuss network delay as a key factor.\\ 

\citet{wang2020performance}. & This article goes into detail about the performance of \ac{hlf} to reveal the varying performances of different ordering services. For this purpose, a network with 20 machines is used, and the different phases of the transaction flow and endorsement policies are considered.\\ 
 
\citet{capocasale2021blockchain}. & The authors of this study present a preliminary performance evaluation of \ac{hlf} v2.2 and compare it with the performance of Hyperledger Sawtooth. They conclude that \ac{hlf}’s throughput for non-sequential workloads is considerably better than that of Hyperledger Sawtooth.\\

\citet{sedlmeir2021dlps}. & This article presents the \ac{dlps} benchmarking framework as an alternative to the widely-used Caliper test suite. The authors evaluate the performance of three different \ac{hlf} networks and compare it to that of other blockchain implementations.\\

\citet{thakkar2021scaling}. & This paper examines the performance of \ac{hlf} v1.4 with regard to horizontal scaling (e.g., by adding more nodes) and vertical scaling (e.g., by varying the number of CPUs per node). Based on these observations, the authors propose an optimization of the \ac{hlf} architecture, including pipelined execution of validation and commit phase.\\ 

\citet{toumia2021evaluating}. & For this performance evaluation of \ac{hlf}, the authors use the Caliper benchmarking framework as well as the fabcar chaincode. The evaluation focuses on the comparison of single ordering service with multi ordering service and considers mixed workloads.\\

\citet{xu2021}. & The authors of this study developed a theoretical analysis framework to study the performance of \ac{hlf} under special consideration of the execute-order-validate logic in \ac{hlf}~v1.4. By means of a series of experiments, they compare the results with simulations to verify the theoretical model.\\

\end{tabular}
}
\vspace{5pt}
\caption{An annotated bibliography of the literature on performance investigations of \ac{hlf}.} 
\label{table:literature_review_measurements}
\end{table}

The study of~\citet{pongnumkul2017performance} marks the first thorough performance analysis of \ac{hlf}. By comparing the Go implementation of the Ethereum client (Geth) to \ac{hlf}, the authors demonstrated the potential performance benefits of using a private permissioned blockchain. 
The following year,~\citet{dinh2018untangling} standardized a performance analysis of private permissioned blockchains by introducing the first systematic benchmarking framework: Blockbench~\citep{BlockbenchRepo}. Blockbench makes heavy use of Yahoo! Cloud Serving Benchmark and Smallbank, both of which are benchmarking frameworks for conventional IT systems with a focus on centralized databases. The authors also compared \ac{hlf} to Geth and Parity. Rather than opt for the re-architectured v1.0 of \ac{hlf},~\citet{dinh2018untangling} used v0.6 for their comparison as they gained far better performance results with the older release.

Subsequent work has made almost exclusive use of \ac{hlf} $\geq$v1.0. Compared to the findings of~\citet{dinh2018untangling}, who accomplished approximately 1,000 transactions per second,~\citet{androulaki2018hyperledger} attained far higher performance statistics with the newly introduced architecture of v1.x. Having analyzed the preview version of v1.1 in detail, they demonstrated that \ac{hlf} has the potential to cope with over 100 peers and, in the right circumstances, perform more than 3,500 transactions per second. However, since the performance results of~\citet{baliga2018performance} were significantly lower than those of~\citet{androulaki2018hyperledger}, it is clear that the potential performance of \ac{hlf} is contingent on various factors, such as its benchmarking framework, its release version, and the employed hardware. Therefore, later research extended testing to multiple parameters and newer release versions of \ac{hlf}. For instance,~\citet{thakkar2021scaling} and~\citet{kuzlu2019performance} analyzed v1.4 of \ac{hlf}, and both studies lend further credence to the complexity of performance tests of blockchain systems. In particular,~\citet{kuzlu2019performance} concluded that it is not merely the specific infrastructure on which the blockchain resides that has a major impact on performance, but also the design of the transactions and thus their type and number. More recent studies have focused on \ac{hlf} v2.0. In particular,~\citet{dreyer2020} have conducted the first measurements of the performance of \ac{hlf} v2.0, indicating that it has improved significantly in comparison to older versions of the blockchain framework. More recently still, \citet{toumia2021evaluating} have evaluated the performance of \ac{hlf} v2.2. Unfortunately, neither of these studies presents in-depth and comparable results for v1.x and v2.x, since the trial runs were not conducted under the same conditions.

So, although later work introduced further influencing factors, the results of~\citet{androulaki2018hyperledger} and~\citet{Thakkar:2018:Fabric} remain the more complete presentations. Table~\ref{table:satisfaction} demonstrates that later work focuses primarily on specific characteristics, such as a sole analysis of the effect of very high network delays. It is important to note, however, that a wide range of other factors has a similarly significant impact on performance, including different benchmarking tools and definitions of key metrics~\citep{sedlmeir2021dlps}, which is why such highly focused studies can only indicate certain trends and first insights into partial developments but are difficult to integrate into the results of other researchers.

\begin{table}[!tb]
\centering
\renewcommand{\arraystretch}{1.25}
\resizebox{\columnwidth}{!}{
\begin{tabular}{l|P{2cm}|P{1.5cm}|P{2cm}|P{1.75cm}|P{1.5cm}|P{2cm}|P{1.5cm}|P{1.5cm}}
  
\textbf{~Source} & \textbf{Fabric version} & \textbf{Vertical scaling} & \textbf{Horizontal\newline scaling} & \textbf{Database} & \textbf{Private data} & \textbf{Multiple workloads} & \textbf{Network\newline delays} & \textbf{Crashing\newline nodes}\\\midrule
This paper & 2.0 (1.4) & \cmark & \cmark & both & \cmark & \cmark & \cmark & \cmark \\\midrule
  
\citet{pongnumkul2017performance} & 0.6 & \xmark &\xmark & LevelDB &\xmark &\xmark &\xmark& \xmark \\
\citet{androulaki2018hyperledger} & 1.1 & \cmark & \cmark & LevelDB & \xmark & \cmark & \phantom{ $^1$}\cmark $^2$ & \xmark \\
\citet{baliga2018performance} & 1.0 & \xmark &\xmark & LevelDB &\xmark &\cmark &\xmark& \xmark \\
\citet{dinh2018untangling} & 0.6 & \xmark & \cmark & LevelDB & \xmark & \xmark & \xmark & \xmark \\
\citet{hao2018performance} & 1.0 & \xmark &\xmark & n/a &\xmark &\xmark &\xmark& \xmark \\
\citet{nasir2018performance} & 1.0 (0.6) & \xmark &\cmark & both &\xmark &\xmark &\xmark& \xmark \\
\citet{Thakkar:2018:Fabric} & 1.1 & \cmark & \cmark & both & \xmark & \cmark & \xmark & \xmark \\

\citet{koushik2019performance} & n/a $^3$ & \xmark & \cmark & n/a & \xmark & \xmark & \xmark & \xmark \\

\citet{kuzlu2019performance} & 1.4 & \xmark & \xmark & CouchDB & \xmark & \cmark & \xmark & \xmark \\  
\citet{nguyen2019impact} & 1.2 & \xmark & \xmark & n/a & \xmark & \xmark & \phantom{$^2$}\cmark $^1$ & \xmark \\

\citet{dabbagh2020performance} & 1.4 (1.1--1.3) & \xmark & \xmark & n/a & \xmark & \xmark & \xmark & \xmark \\

\citet{dreyer2020} & 2.0 (0.6/1.0) & \xmark & \cmark & n/a & \xmark & \cmark & \xmark & \xmark \\

\citet{geneiatakis2020} & 1.1 & \xmark & \cmark & CouchDB & \xmark & \cmark & \cmark & \xmark \\

\citet{wang2020performance} & 1.4 & \xmark & \cmark & n/a & \xmark & \cmark & \xmark & \xmark \\

\citet{capocasale2021blockchain} & 2.2 & \xmark & \xmark & n/a & \xmark & \cmark & \xmark & \xmark \\

\citet{sedlmeir2021dlps} & 1.4 & \xmark & \cmark & both & \xmark & \xmark & \xmark & \xmark \\

\citet{thakkar2021scaling} & 1.4 & \cmark & \cmark & n/a & \xmark & \cmark & \xmark& \xmark \\

\citet{toumia2021evaluating} & 2.2 & \xmark & \cmark & CouchDB & \xmark & \cmark & \xmark & \xmark \\

\citet{xu2021} & 1.4 & \xmark & \cmark & n/a & \xmark & \cmark & \cmark & \xmark \\

\end{tabular}
}
\vspace{1em}~\\
\renewcommand{\arraystretch}{1.2}
\resizebox{\columnwidth}{!}{
\begin{tabular}{@{}l@{}}
\vspace{-0.5em}
$^1$ The authors only considered delays of 1,000 ms and more, which is far more than the delays that typically occur in a worldwide distributed system.\\
\vspace{-0.5em}
$^2$ The authors only considered network delay in a single setting, without stating the actual delay between the data centers involved and without further analysis \\~~~~of the impact of different delays.\\
\vspace{-0.5em}
$^3$ The authors did not state the exact version of the investigated Fabric SDK. However, based on the description of the system, we assume this to be Hyperledger Fabric $\geq$ v1.0. 
\end{tabular}
}
\vspace{0.5pt}
\caption{Evaluation of the measurements conducted for the research papers discussed in the above bibliography.}
\label{table:satisfaction}
\end{table}

In summary, while the literature published to date has provided some significant first insights into the properties of Fabric, these insights have been partial as the literature’s parameters are still defined in rather narrow terms. It is also important to realize that the results presented to date are only reproducible to a somewhat limited extent because the methodologies by which they were attained have often been accompanied by minimal descriptions. There is, then, considerable room to improve the general validity of these results.

\section{Evaluation Framework}
\label{sec:methods}

To advance the current understanding of private permissioned blockchains and conduct further analysis of the multiple variables that may impact the performance of \ac{hlf}, we opted for standardized benchmarking. Having examined the many tools used for blockchain benchmarking in the literature we analyzed, we found that those best suited to \ac{hlf} are Blockbench~\citep{BlockbenchRepo}, Caliper~\citep{Caliper}, and the \ac{dlps}~\citep{sedlmeir2021dlps}. 
Blockbench and Caliper, however, do not adequately speficy how they define and determine the key performance metrics, particularly throughput and latency. Since the algorithm by which these are determined remains unclear, we opted for the open-source framework \ac{dlps}. This had the added advantage that \ac{dlps} allows for sophisticated network deployment with the use of cloud services, which enabled us to test an unprecedented range of configurations.

Our benchmarking covers all the variables identified in the above review process (see Table~\ref{table:satisfaction}). By conducting tests that went beyond those prior studies, we identified seven additional variables with the potential to impact the performance of \ac{hlf}. Since the \ac{dlps} did not cover all the particularities of \ac{hlf}, our first order of priority was to upgrade the \ac{hlf} version supported by DLPS to \ac{hlf} 2.0 and include multi-channel setups. Our second extension was to add support for private transactions and complex queries. By way of a third amendment, we extended the supported architectural parameters in such a way as to allow the CouchDB and ordering node docker containers to run either on the same node as the peers or on separate nodes. In doing so, we accounted for the possibility that splitting tasks on multiple machines or joining them to reduce cross-instance latencies might help to increase performance.
Our fourth change was to add support not only for simulating network delays but also for multi-datacenter deployments. Our fifth and final improvement was to refine the overall benchmarking process, evaluate single-core CPU usage, analyze traffic stats, and add capabilities to trigger automatic crashes of orderers and peers. To name but one example, this required the dynamic identification of the current leader in the RAFT ordering service. Consequently, the final framework allows testing for all previously mentioned variables found in prior research and extends them with new unique features that, according to our literature review, had not been investigated to date. Table~\ref{tab:requirements} provides a description of all variables considered for this benchmarking. With the publication of this paper, we make our improvements to \ac{dlps}, as well as the configurations and results of all the experiments we conducted for this study, available on the \ac{dlps} GitHub repository~\citep{DLPS}.

\begin{table}[!htb]
\centering
\resizebox{0.85\columnwidth}{!}{
\begin{tabular}{L{2.5cm}|L{4cm}|p{11.5cm}}
\midrule
\bf{Group} & \bf{Design choices} & \bf{Answered questions} \\\midrule
\multirow{4}{*}{\shortstack[l]{Architecture\\ (Sec.~\ref{subsec:architecture})}}
& Number of organizations, peers, and orderers & How does an organization's configuration of its peers and orderers impact the performance of \ac{hlf}? \\
& Endorsement policy & What is the impact of different endorsement policies? \\
& Number of channels & Does changing the number of channels impact performance? \\
& Database location & Does the separation of the database from the peer core functions improve the performance of \ac{hlf}? \\
\midrule
\multirow{3}{*}{\shortstack[l]{Setup\\(Sec.~\ref{subsec:setup})}}
& Hardware & What is the impact of different computer specifications, particularly CPU? \\
& Database type & How do different databases for the ledger state, such as CouchDB or LevelDB, impact the performance of \ac{hlf}?\\
& Block parameters & How does the choice of blocksize and blocktime affect performance? \\
\midrule
\multirow{4}{*}{{\parbox[c]{2cm}{\shortstack[l]{Business Logic\\(Sec.~\ref{subsec:business_logic})}}}}
& Private data & How does using private transactions impact the performance of \ac{hlf}? \\
& I/O-heavy workload & How do transactions that trigger I/O-heavy chaincode impact the performance
of \ac{hlf}? \\
& CPU-heavy workload & How do transactions that trigger CPU-heavy chaincode impact the performance of \ac{hlf}? \\
& Reading vs. writing & What are the essential differences between read and write performance?\\
\midrule
\multirow{2}{*}{\shortstack[l]{\phantom{H}\\Network\\(Sec.~\ref{subsec:network})}}
& Delays & To what extent do network delays impact performance? \\
& Bandwidth & What are the bandwidth requirements for different architectures and throughputs?\\ 
\midrule
\multirow{4}{*}{\shortstack[l]{Robustness\\(Sec.~\ref{subsec:robustness})\\\phantom{H}\\\phantom{H}}}
& Node crashes & How do crashing nodes impact the performance of Hyperledger Fabric? \\
& Temporal distribution of requests & How do changes in the temporal request distribution affect the performance of \ac{hlf}?
\end{tabular}
}
\vspace{5pt}
\caption{Design choices and network specifics in need of thorough analysis.}
\label{tab:requirements}
\end{table}

\begin{figure}[!tb]
    \centering
    \includegraphics[width=0.95\textwidth, trim=0cm 0cm 0cm 0cm, page=1, clip]{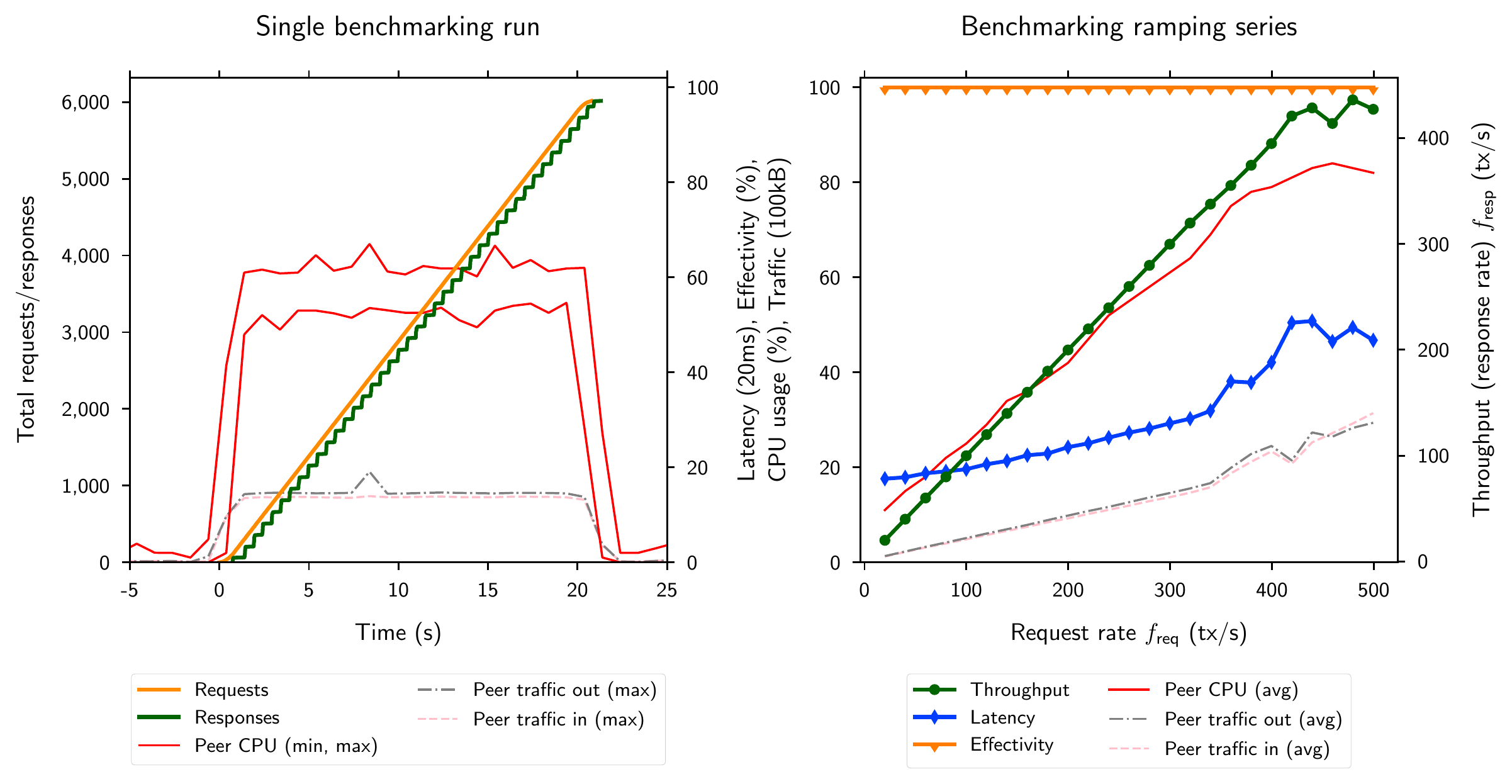}
    \caption{Exemplary benchmark run charts.}
    \label{fig:method_chart}
\end{figure}

We performed the testing in an incremental manner to ensure the reliability of our results. The left chart in Figure \ref{fig:method_chart} describes a single benchmarking run. We used a series of these runs to create a benchmarking ramping series (see the right chart in Figure~\ref{fig:method_chart}).
We also created a configuration file that specifies all particularities of the \ac{hlf} network. The \ac{dlps} uses this file and automatically sets up a blockchain and client network in \ac{aws} prior to the benchmarking process. A single benchmarking run in the \ac{dlps} comprises the sending of requests from clients to the network for a specific duration at a specific rate $f_{\text{req}}$, namely the slope of the requests (see orange requests curve). The receipt of confirmations that the transactions have been processed successfully is illustrated by a response curve (see green response curve). In this curve, one can see how distinguished blocks result in steps as they mark a quasi-simultaneous confirmation of the included transactions. As the graphic shows, the linear response regression describes a curve with a slope that corresponds to the average rate of responses. The average time between sending and receiving confirmation of a specific transaction marks the latency. Likewise, as long as the linear regressions remain parallel, the distance between where the regression for the request and response curves intersect with the $x$-axis marks the latency.

By starting at a low request rate and repeating tests at an increased request rate in case the network can process requests at the given rate (see $x$-axis of right Chart in Figure \ref{fig:method_chart}), we can localize the maximum throughput, where another increase in the request rate does not improve the response rate any further or indeed deteriorates it due to queueing or overstress. In Figure~\ref{fig:method_chart}, this behavior can be observed in the right chart at a request rate of approximately~450~tx/s. In each case, we monitored the effectivity -- the rate of transactions that were successfully operated -- along with controlled resource stats, such as CPU usage and network traffic, to gather any additional information that might help to locate the bottleneck. More information about the \ac{dlps} can be found on GitHub~\citep{DLPS} and in the associated paper~\citep{sedlmeir2021dlps}.

By default, the deployments and tests done with the \ac{dlps} are highly homogeneous and symmetric. We associate each client with one blockchain node, while each blockchain node is associated with the same number, typically larger than one, of clients. This makes it possible to send requests completely uniformly, which is to say that at $f$~transactions per second, one transaction is sent each $\tfrac{1}{f}$ second, which is again uniformly distributed among the clients. For example, if we have a 10-node network and 20~clients, and a request rate of 100\,tx/s, every client sends requests at 5\,tx/s. Moreover, we ensure a uniform offset between the clients. In case a client has multiple cores, and we use multiple workers for multi-threading, we make sure that there is a homogeneous offset, too. It is worth noting that at high request rates, the offset is harder to enforce and far less relevant. Meanwhile, a high degree of uniformity is relevant if one is to measure maximum throughput correctly when it is low, as only then there are no spikes in the nodes' workload.

We used instances from the m5 series in AWS because they strike a good balance between computation, networking, and disk operations, all of which are necessary for blockchain nodes. Table~\ref{tab:instances} features details on these instances. The fully automatic setup with the \ac{dlps} takes a minimum of 10~minutes, a reasonable test approximately an hour. We found that reducing the duration of a test increased the variance of its results and tended to overestimate the performance in our tests. Generally speaking, any modifications of network parameters require the blockchain to be completely restarted. 

Bearing this in mind, we decided to use a small network (four organizations, each with two peers, one orderer, and four clients) with AWS m5.large instances as our default parameters. We also decided to vary different individual or small subsets of parameters, starting from this default, to keep costs and time within reasonable bounds. Figure~\ref{fig:default} illustrates the most significant of the remaining default parameters for the \ac{hlf} architecture, while Figure~\ref{fig:default2} gives an overview of the benchmarking settings.
In total, then, our default configuration comprises eight peers and four orderers, as well as 16 clients, in a one-channel network with RAFT consensus. At the start of our experiments, the latest Fabric version was v2.0, which is why we conducted all experiments with this version. When v2.2 was released, we also made spot checks for the purpose of validation and cross-referencing but noticed no significant performance changes. The remaining parameters are described in detail in the dedicated \ac{dlps} repository~\citep{DLPS}.

\begin{table}[!htb]
\begin{center}
\resizebox{0.7\columnwidth}{!}{
\begin{tabular}{c|c|c|c|c}
\textbf{Name} & \textbf{vCPUs} &	\textbf{Memory (GiB)} &	\textbf{Network (Gbps)} & \textbf{Storage (Mbps)} \\ \midrule
m5.large & 2 & 8 & Up to 10 & Up to 4,750 \\
m5.xlarge &	4 & 16  & Up to 10 & Up to 4,750 \\
m5.2xlarge & 8 & 32 & Up to 10 & Up to 4,750 \\
m5.4xlarge & 16 & 64 & Up to 10 & 4,750 \\
\end{tabular}
}
\vspace{5pt}
\caption{\label{tab:instances}Used instance types in the AWS m5 series, all based on Intel Xeon\textsuperscript{\tiny\textregistered} Platinum 8175M processors (up to 3.1 GHz). We added 16\,GB of SSD storage. As the operating system, we used Ubuntu 18.04 LTS. Source: \citep{aws}}
\end{center}
\end{table}

\begin{figure}[!htb]
\noindent\begin{minipage}[t]{0.35\textwidth}
\begin{lstlisting}[language=json,numbers=none,frame=tlrb]
{
  "node_type": "m5.large",
  "fabric_version": "2.0.0",
  "fabric_ca_version": "1.4.4",
  "thirdparty_version": "0.4.18",
  "channel_count": 1,
  "database": "CouchDB/LevelDB",
  "external_database": "False",
  "internal_orderer": "False",
  "org_count": 4,
  "peer_count": 2,
  "orderer_type": "RAFT",
  "orderer_count": 4,
  "batch_timeout": 0.5,
  "max_message_count": 1000,
  "absolute_max_bytes": 10,
  "preferred_max_bytes": 4096,
  "tls_enabled": "True",
  "endorsement": "OutOf(2, 4)",
  "private_fors": 2,
  "log_level": "Warning",
  "client_type": "m5.large",
  "client_count": 4,
}
\end{lstlisting}
\caption{Default settings for the \ac{hlf} network architecture.}
\label{fig:default}
\end{minipage}
\noindent\begin{minipage}[t]{0.15\textwidth}
~
\end{minipage}
\noindent\begin{minipage}[t]{0.35\textwidth}
\begin{lstlisting}[language=json,numbers=none,frame=tlrb]
{
  
  "duration": 20,
  "localization_runs":2,
  "repetition_runs": 0,
  "method": "writeData",
  "mode": "public",
  "shape": "smooth",
  "delay": 0,
  "r2_bound": 0.9,
  "frequency_bound": 100,
  "latency_bound": 10000,
  "delta_send": 0.5,
  "delta_receive": 0.5,
  "success_bound": 0.8,
  "retry_limit": 2,
  "ramp_bound": 2,
  "success_base_rate": 0.8,
  "success_step_rate": 0.04,
  "failure_base_rate": 0.8,
  "failure_step_rate": 0.04,
  "delta_max_time": 10
  
}
\end{lstlisting}
\caption{Default settings for the benchmarking logic.}
\label{fig:default2}
\end{minipage}
\end{figure}

In total, our experiments involved near enough 2,000 hours of testing, setting up approximately 1,500 \ac{hlf} networks with a total of around 20,000 nodes and 40,000 clients, and sending more than 200 million transactions. As part of this process, we also collected 100\,GB of log files to record such factors as the send and response times of each transaction as well as multiple resource stats such as CPU, memory, disk usage, ping, and traffic for each node and client.

\section{Benchmarking Results}
\label{sec:results}

\subsection{Stability of the default setup and comparison of software versions and databases}
\label{subsec:first_observations}

Our first comparative analysis of the variously modified default architecture focused on relevant parameters we identified in the above literature review and in our experience of working with the \ac{dlps} (see Figure~\ref{fig:various_throughput}).
The error bars and areas in the charts in the following figures, which we created with seaborn~\cite{Waskom2021seaborn}, represent the standard deviation obtained from conducting every experiment three times, and as these error bars indicate, the results are highly consistent and reproducible. Our default workload is a set of ``simple'' transactions which consist of writing a single key-value pair to the blockchain's database, each pair the size of a few bytes.

Confirming the results of~\citet{Thakkar:2018:Fabric}, we found that write throughput for the default setup with LevelDB is around three times the maximum throughput with CouchDB. Furthermore, we were able to extend this result to private transactions (see  Figure~\ref{fig:various_throughput}).
We observed that throughput was impacted only at an insignificant rate by the following modifications: doubling the number of clients so as to distribute client workload to more workers (0\,\% impact on performance with public transactions and 4\,\% increase of performance with private transactions), doubling the number of channels (2\,\% decrease of performance with public transactions and 13\,\% increase of performance with private transactions), deactivating TLS and switching to a centralized (``solo'') orderer (13\,\% increase on performance with public transactions and 3\,\% increase of performance with private transactions). This indicates that the bottleneck of the default architecture in our setup is neither the number of clients and channels nor the ordering service and TLS. 
While the results of~\citet{Thakkar:2018:Fabric} lend further credence to our conclusion that the ordering service is not a bottleneck in a similar architecture, they find that doubling the number of channels considerably increases CPU utilization and with it throughput. In our case, however, CPU utilization by peers is already very close to the maximum, and this is true on all virtual cores with one channel. This observation indicates that the two-channel configuration does not, ultimately, exhibit a higher throughput. In our case, the difference between single-channel and dual-channel setup was small, with a variation of only 9\,\% for private transactions. Public transactions did not even show any significant impact. It is worth noticing, however, that these numbers represent the results with CouchDB. With LevelDB, the relative deviations tended to be even smaller.

Performance benchmarks with older versions of~\ac{hlf}, particularly those set by~\citet{pongnumkul2017performance}, \citet{Blockbench:2017:Dinh} and \citet{nasir2018performance}, generally yield lower throughput (few hundred tx/s with LevelDB) on considerably better hardware, indicating that the evolution of \ac{hlf} has already led to considerable performance improvements. It came as a surprise, therefore, that we noticed a slight decrease in the performance of v2.0, compared to that of the previous version~1.4.4 for CouchDB, as opposed to the results of~\citet{dreyer2020}. Indeed,~v1.4.4 using CouchDB was about 26\,\% faster with public transactions and 68\,\% faster with private transactions than v2.0. With LevelDB, the difference for private transactions dropped to a mere~11\,\%, and with public transactions~v2.0 was 5\,\% faster than~v1.4.4. We put this discrepancy between our results and those of~\citet{dreyer2020} down to how they arrived at their conclusion since they compared their results for v2.0 with the results of~\citet{nasir2018performance} for v0.6 and v1.4. However, the testing environment of their studies was different, and \citet{dreyer2020} used stronger machines with more computing power, which is most likely why they measured a better performance with regard to v2.0. In contrast, our comparison of v1.4.4 and v2.0 was conducted under otherwise identical conditions.

\begin{figure}[!tb]
    \centering
    \includegraphics[width=0.8\linewidth, trim= 0cm 0cm 0cm 0cm ]{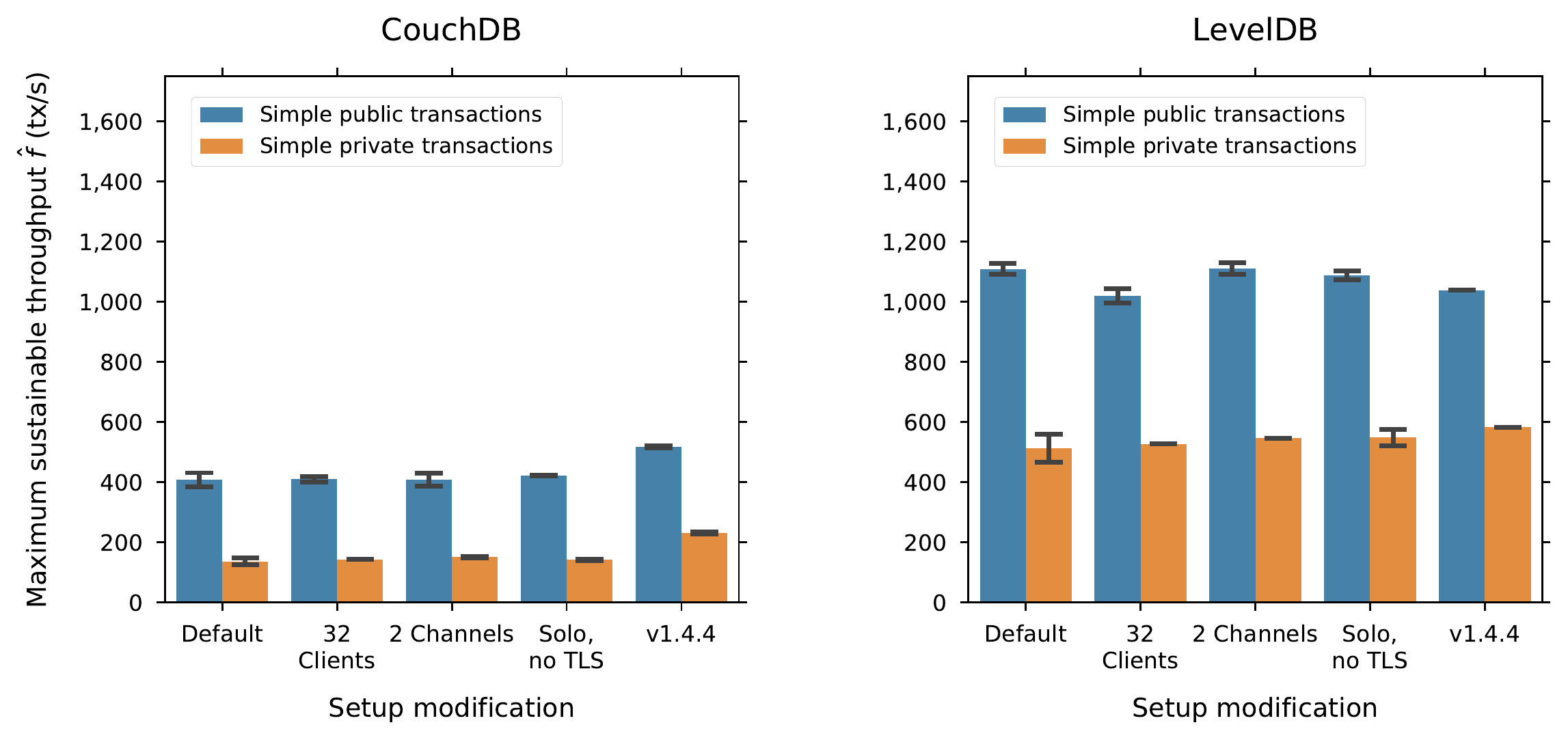}
    \caption{A side-by-side comparison of different architectures in comparison; the configuration for the default setup is described at the end of Section~\ref{sec:methods}.} 
    \label{fig:various_throughput}
\end{figure}

\subsection{Architecture}
\label{subsec:architecture}

\subsubsection{Endorsement Policy}

\begin{figure}[!tb]
    \centering
    \includegraphics[width=0.8\linewidth, trim= 0cm 0cm 0cm 0cm ]{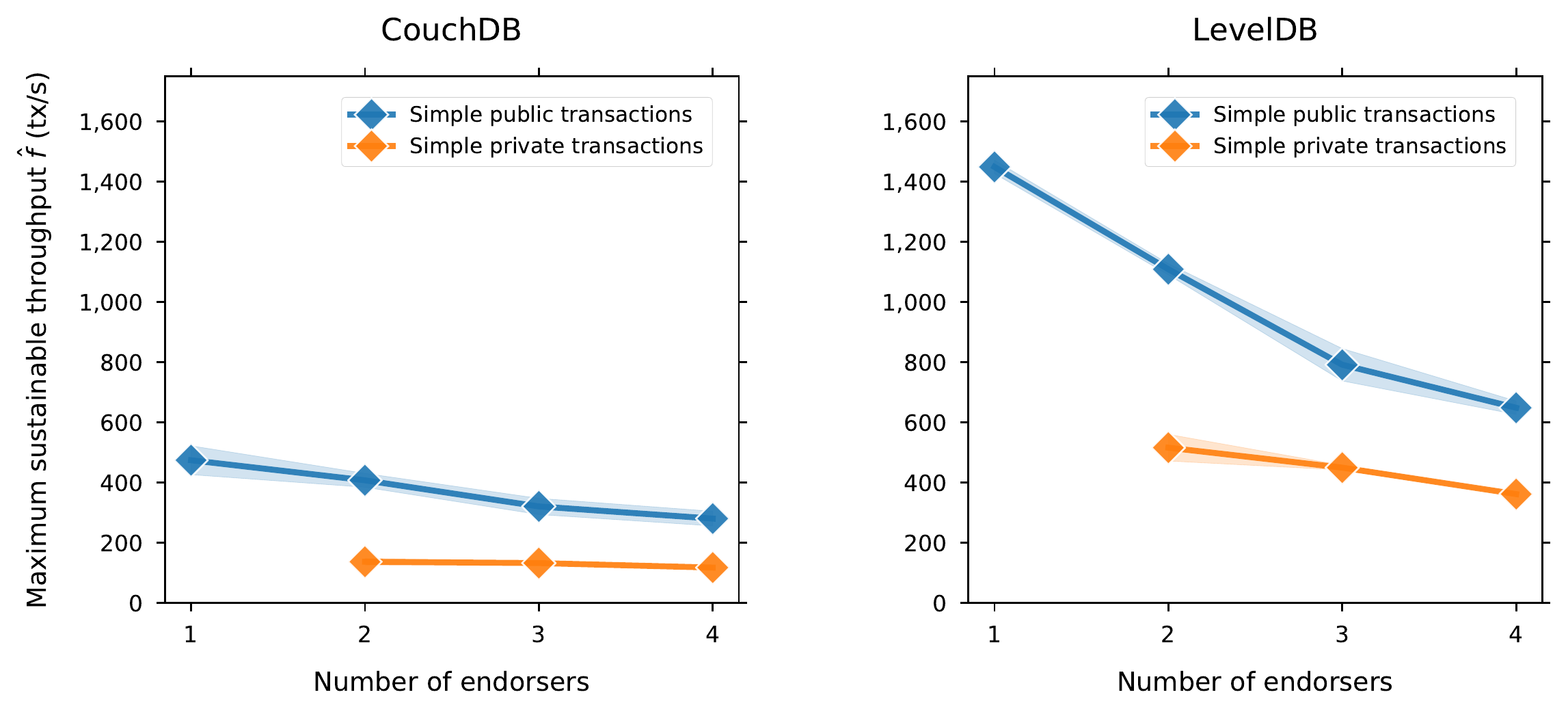}
    \caption{Varying the endorsement policy.}
    \label{fig:endorsement}
\end{figure}

The endorsement policy, described in Section~\ref{sec:background}, is a key setting as it drastically changes the level of redundancy in simulation (execution). For instance, with a rise in the number of endorsers, the overhead increases notably, but so too does the robustness. As illustrated in Figure~\ref{fig:endorsement}, an increase in the number of endorsers leads to the expected corresponding decrease in throughput. In absolute and relative numbers, LevelDB suffers from a much higher performance decrease with a higher number of orderers compared to CouchDB. For example, maximum throughput for simple public transactions with LevelDB decreases by 24\,\% respectively 54\,\% when switching from only one endorser and, thus no cross-checks of correct chaincode execution, to two or four endorsements, respectively. For CouchDB, degradation is 14\,\% respectively 41\,\%.

For private transactions, we looked at pairwise private collections, which is to say private transactions between two organizations. We found that moving from two to four endorsers results in a loss of 14\,\% (CouchDB) and 30\,\% (LevelDB) in maximum throughput. These numbers are notably lower than those attained with public transactions (31\,\% and 42\,\%). Accordingly, when more endorsements are necessary, performance decreases more for LevelDB, both in relative and absolute terms. Surprisingly, however, one endorser of private transactions proved to be an outlier of sorts. Indeed, we noticed a somewhat strange behavior of Fabric which resulted in throughput in the one-digit range as soon as multiple clients were requesting transactions from different peers. So far, however, we have not been able to determine the reason for this anomaly.

\subsubsection{Network Architecture}

\begin{figure}[!tb]
    \centering
    \includegraphics[width=0.8\linewidth, trim= 0cm 0cm 0cm 0cm ]{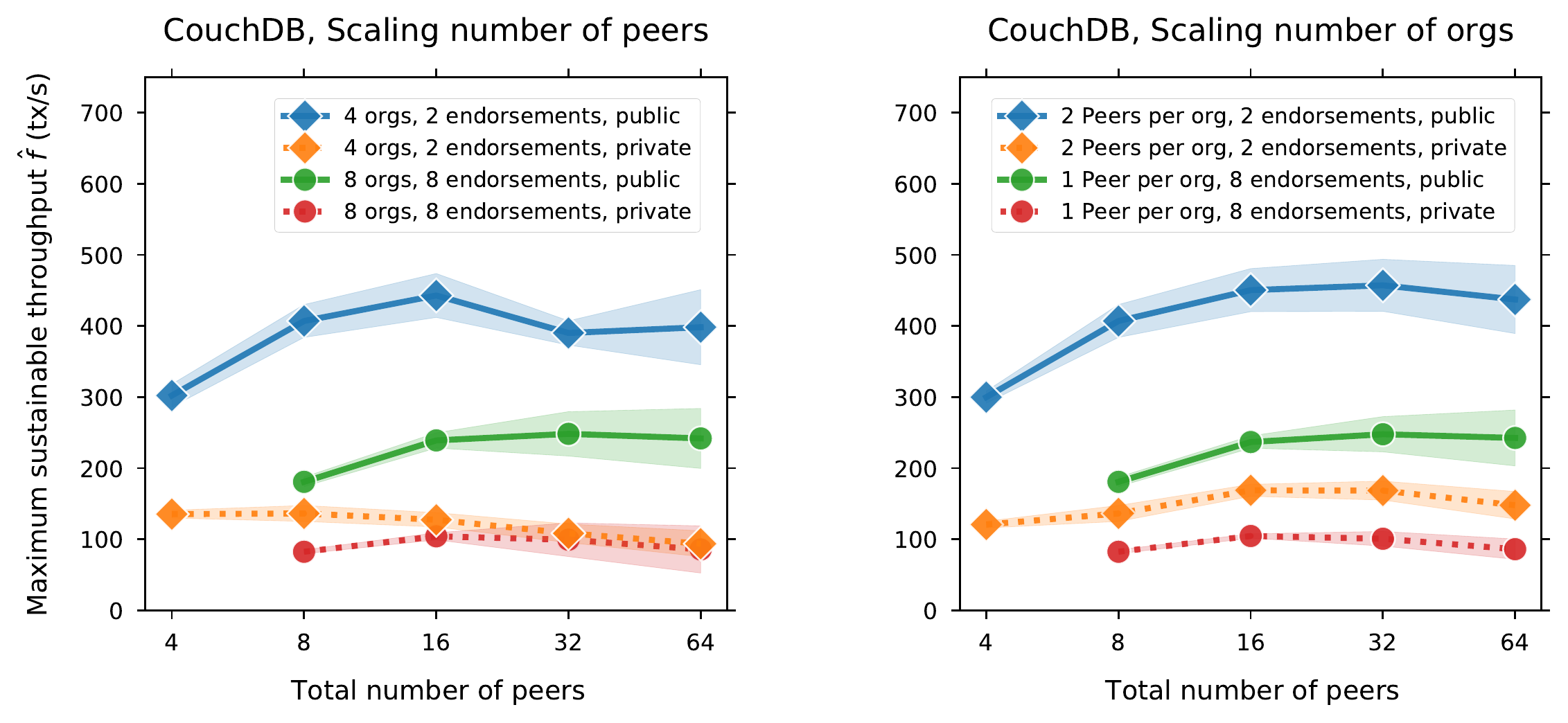}
    \includegraphics[width=0.8\linewidth, trim= 0cm 0cm 0cm 0cm]{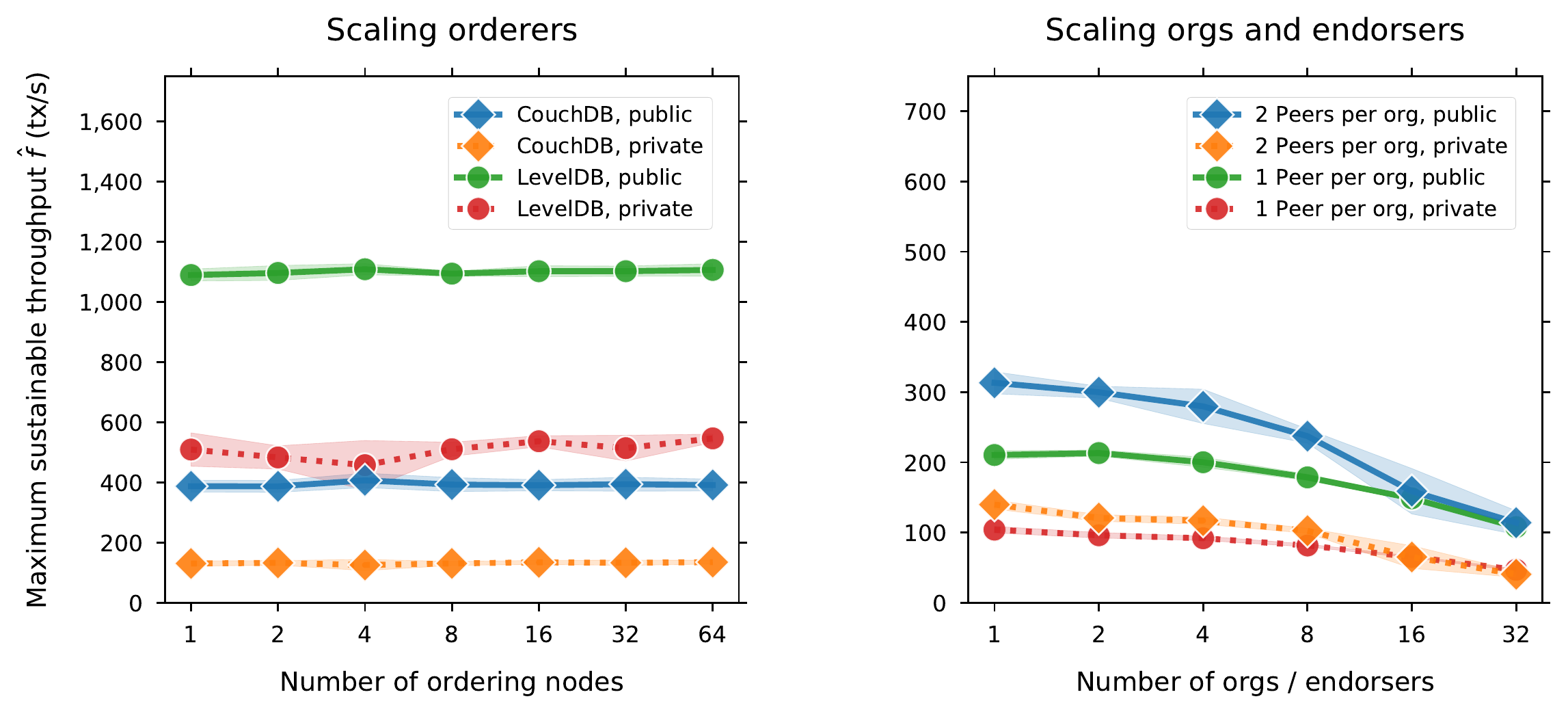}
    \caption{Different scalability parameters in comparison.}
    \label{fig:scalability_nodes}
\end{figure}

Initially, we see an increase in maximum throughput when increasing the number of peers per organization while keeping the number of endorsers constant (see Figure~\ref{fig:scalability_nodes}). Likewise, increasing the number of organizations while keeping the number of peers per organization and the number of endorsers constant increases maximum throughput. However, we also notice that maximum throughput decreases again for large network sizes, so there seems to be an optimum. For the given setup, this optimum is at eight peers per organization and an endorsement policy of two out of eight. We could, therefore, improve public transaction performance by up to 32\,\% by adjusting the number of peers. With private transactions, the improvements are a little less impressive yet still significant at a 21\,\% performance increase with eight peers per organization rather than with two peers per organization.

Scaling the number of organizations and the endorsement policy up proportionately only slightly reduces throughput for smaller networks, a potential reason being that the endorsement workload for each peer remains constant and other operations like networking and committing are not the bottleneck in this regime. Nevertheless, for larger network sizes, throughput degrades considerably, and we also see that the difference between having one and two peers per organization on throughput becomes negligible. This makes sense, as networking becomes the bottleneck in this regime, and adding further peers inside an organization only further reduces the (insignificant) endorsement workload for each peer. Moreover, it increases the already significant networking effort for the anchor peer who receives blocks from the ordering service and distributes them further to the other peers in the associated organization.

For scaling the number of RAFT orderers, we expected a performance decrease but none was yet to be observed in our chosen scenarios. It would appear that, below a request rate of~1,500\,tx/s, the ordering service is not a bottleneck for up to 64 orderers, althought it might become a bottleneck for larger ordering services. Using a RAFT ordering service with up to 64 nodes should be sufficient in practically any scenario since this would allow a total of 31 crashes and still ensure the network's functionality.

\subsubsection{Database Location}

\begin{figure}[!tb]
    \centering
    \includegraphics[width=0.8\linewidth, trim= 0cm 0cm 0cm 0cm ]{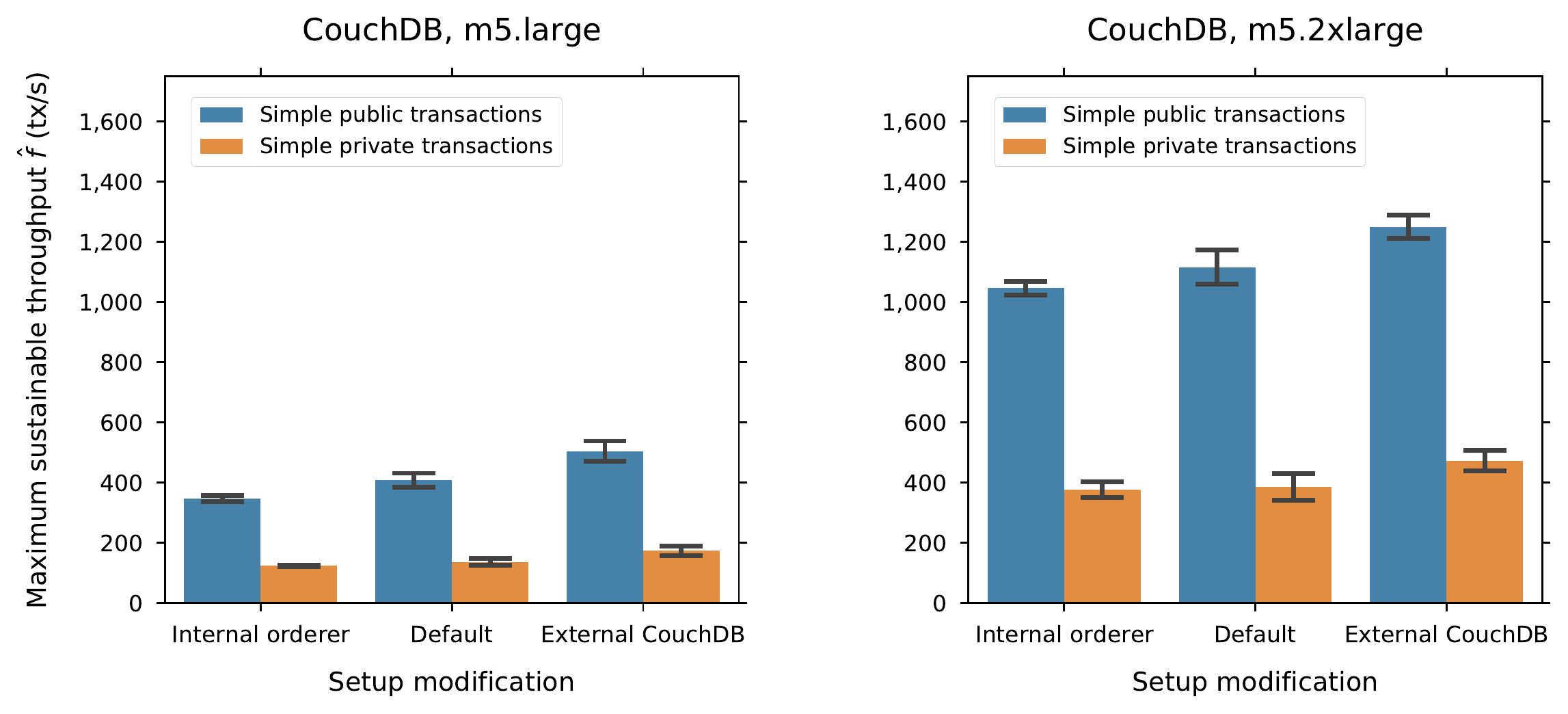}
    \caption{The effect of separating the ordering nodes and the database for CouchDB.}
    \label{fig:scalability_endorsement}
\end{figure}

Deploying databases, orderers, and peers to separate systems facilitates a small boost in performance (see Figure~\ref{fig:scalability_endorsement}). In our default scenario, the database runs on the peer node (which is obligatory for LevelDB) while orderers work on separate nodes. Our results indicate that running both the orderer and the peer on the same node causes only a slight decrease in performance. Disregarding other important factors, separating an organization's \ac{hlf} components on several servers is notably less efficient. In particular, we observed a decrease of only 15\,\% in the case with the m5.large machines and 6\,\% in the case of the m5.2xlarge machines. Meanwhile, running the CouchDB on a separate node has a considerable impact on weaker hardware -- an increase of 23\,\% with m5.large. The throughput improvement is similar for private transactions on m5.large and m5.2xlarge hardware, amounting to 22\,\% respectively 18\,\%, but it becomes smaller (in relative terms) as soon as better equipment is used (an increase of 12\,\% with m5.2xlarge). 

\subsection{Setup}
\label{subsec:setup}

\subsubsection{Database Type}

Early on during our experiments, we realized that \ac{hlf}'s performance is contingent on the choice of database. On average, throughput was two to three times higher with LevelDB than with CouchDB. We also conducted individual measurements for LevelDB and CouchDB on the hardware of our default setup and noticed that for writing a single key-value pair, LevelDB has a throughput of more than~5,000\,tx/s, while a standalone CouchDB manages only around~500\,tx/s. This suggests that in both cases, the databases alone are not the bottleneck, but the individual inefficiencies of those databases have a considerable performance impact. Moreover, CouchDB runs in an individual docker container, whereas LevelDB is integrated into the peer's docker container, which is why the interaction may contribute to the performance differences.

For private data, the difference between the database types tends to be even more noticeable. For example, with private data and m5.large machines, \ac{hlf} was 272\,\% faster with LevelDB than with CouchDB in terms of throughput in the default setup. This confirmed our intuition because a private transaction implies additional write transactions on peers that participate in a private transaction: The payload hash is distributed to all peers in the network, so this involves as many write transaction on each peer as a ``normal'' transaction would. However, legitimate peers also query the private data from the endorsing peers and add them to their database in an additional write transaction (see also the description of the private transaction data flow in Section~\ref{subsec:private_flow}).

\subsubsection{Hardware}
\begin{figure}[!tb]
    \centering
    \includegraphics[width=0.8\linewidth, trim= 0cm 0cm 0cm 0cm ]{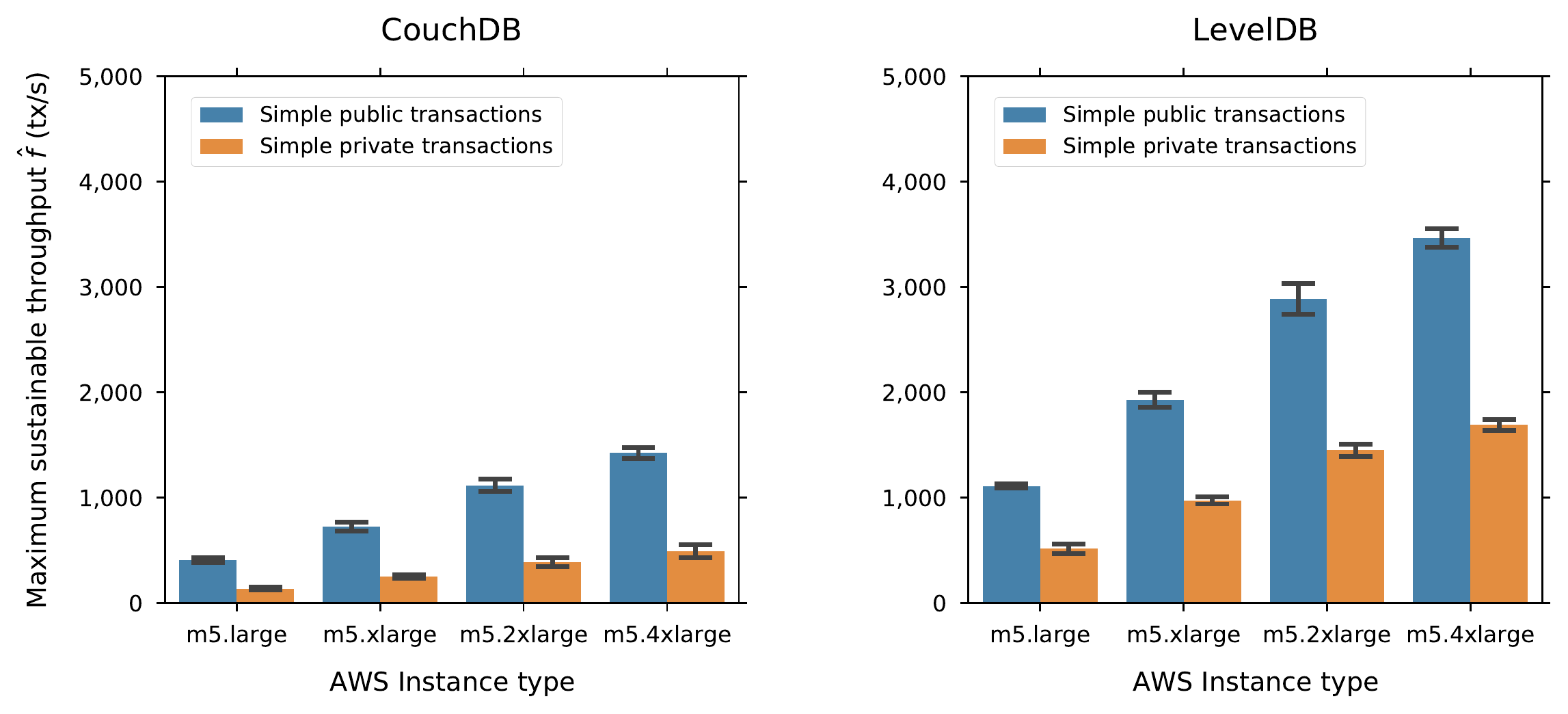}
    \caption{A comparison of different instance types for simple public and private transactions with CouchDB and LevelDB.}
    \label{fig:instance_type_throughput}
\end{figure}

We found it to be important to determine the correlation between machine strength and performance since systems should scale with better hardware (and better network). As long as there are only a few vCPUs, an increase in their number improves performance notably (see Figure~\ref{fig:instance_type_throughput}). For example, when moving from m5.large to m5.xlarge instances, the performance increase for private transactions with CouchDB is 97\,\% and 62\,\% when moving from m5.xlarge to m5.2xlarge instances. Similar margins can be observed for both CouchDB and LevelDB and both public and private transactions. However, the improvement made by moving from m5.2xlarge (8~vCPUs) to m5.4xlarge (16~vCPUs) is rather small (less than 25\,\% for CouchDB and less than 20\,\% for LevelDB for both public and private transactions), particularly when one takes into account that this also involves twice the costs for hardware and cloud services. When we took measurements with m5.8xlarge instances, we noted that performance improvements were even lower than for moving from m5.2xlarge to m5.4xlarge. Besides, crashes of peers became quite frequent, particularly for LevelDB, which led to our maximum throughput results to be even worse than those for m5.4xlarge instances. Identifying the reasons for this behavior promises to be an interesting starting point for future improvements of \ac{hlf} and should allow for better scaling with hardware.

\begin{figure}[!tb]
    \centering
    \includegraphics[width=0.8\linewidth]{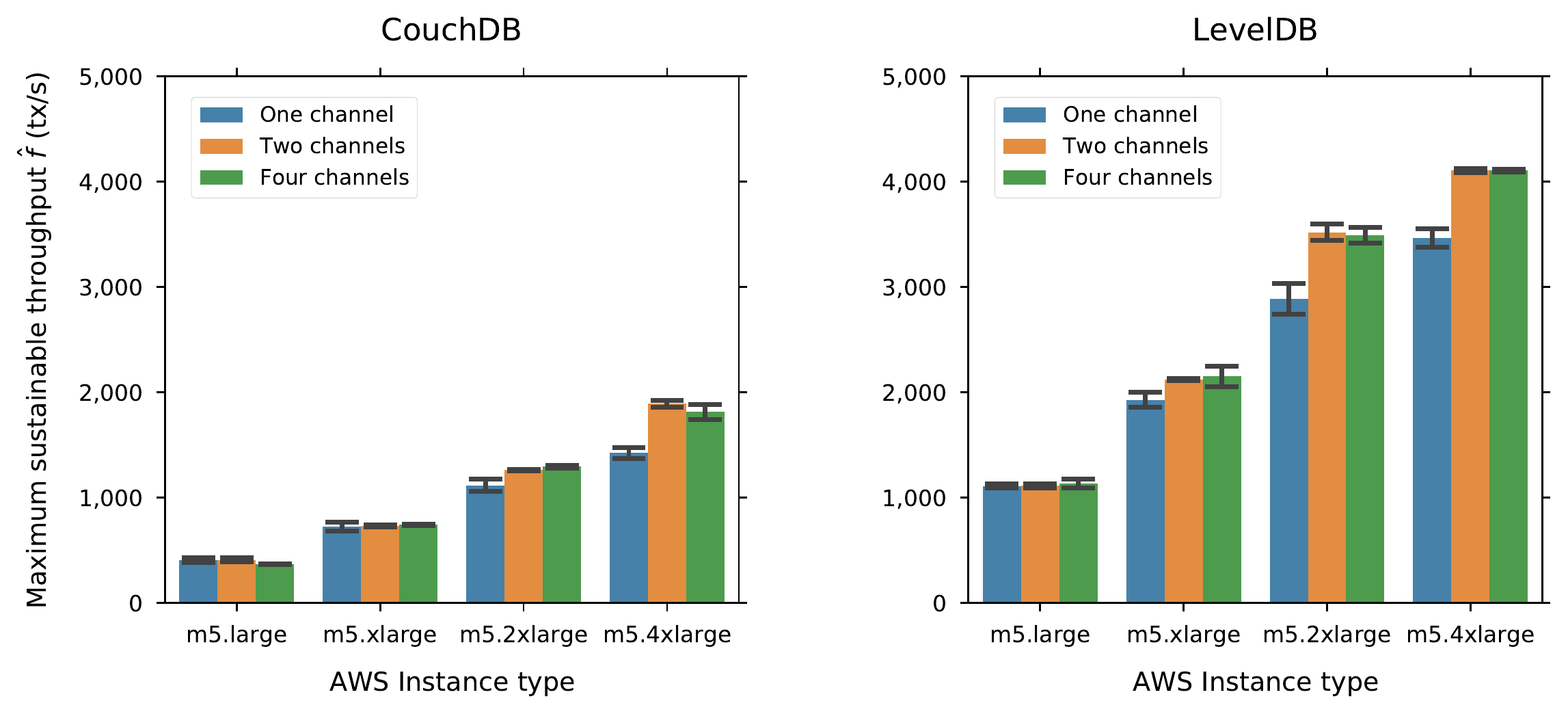}
    \caption{Using multiple channels with varying hardware for simple public transactions with CouchDB and LevelDB.}
    \label{fig:hardware_channels}
\end{figure}

Like~\citet{thakkar2021scaling}, we also observed that CPU utilization drops for hardware with a high number of cores. \citet{thakkar2021scaling} further argue that throughput can be increased by using more peers on multiple channels. However, this is basically the same as running multiple blockchains instead of one, but at present only cross-chain read-operations between the blockchains (channels) are supported. Furthermore, our experiments suggested that for hardware with many cores, the CPUs cannot be fully utilized, and there is not a single core that reaches more than 90\,\% CPU utilization. Therefore, the computational tasks seem well parallelized and suggest that, ultimately, writing to disk during the commit phase may be the bottleneck. Nevertheless, we were keen to ascertain whether using multiple channels could leverage additional resources. As our results indicate, this does, indeed, seem to be the case, but only to a small extent. Our results (Figure~\ref{fig:hardware_channels}) confirm that an increasing the number of channels has a small impact (an average of 12\,\%, regardless of the database type) when switching from a single-channel setup to a dual-channel setup. Adding further channels leads to no noticeable further improvement in maximum throughput.

\subsubsection{Block Parameters}

New blocks are generated by the ordering service when either the maximum blocksize is reached or the time that has passed since the generation of the last block is longer than the blocktime. For the purpose of our experimental setup, we selected a request rate of 500~tx/s, at which we observed that the response rate (throughput) cannot exceed 500~tx/s, yet it will be around 500~tx/s when the blockchain can handle at least 500~tx/s. Since the maximum blocksize is 1,000 transactions, and the blocktime is two seconds, this means that blocks cannot comprise more than 1,000~transactions. Therefore, it will always be blocktime that triggers the creation of a new block. By the same logic, too great a reduction in blocktime results in a throughput decay caused by the block production overhead. Also \citet{Thakkar:2018:Fabric} have noted this positive correlation between block size and maximum throughput. For larger blocktimes, the transaction workload dominates, which is why performance tends to be far less contingent on changes in blocktime. Latency naturally increases with blocktime, as it is always the associated timeout that triggers the creation of new blocks. On the other hand, decreasing maximum blocktime by decreasing latency below 0.5\,s also heavily decreases throughput. Consequently, we find that a block timeout of around 0.5\,s constitutes a sweet spot -- any decrease would make throughput significantly worse, and any increase does not substantially improve throughput yet increases latency. 

\begin{figure}[!tb]
    \centering
    \includegraphics[width=0.8\linewidth, trim= 0cm 0cm 0cm 0cm ]{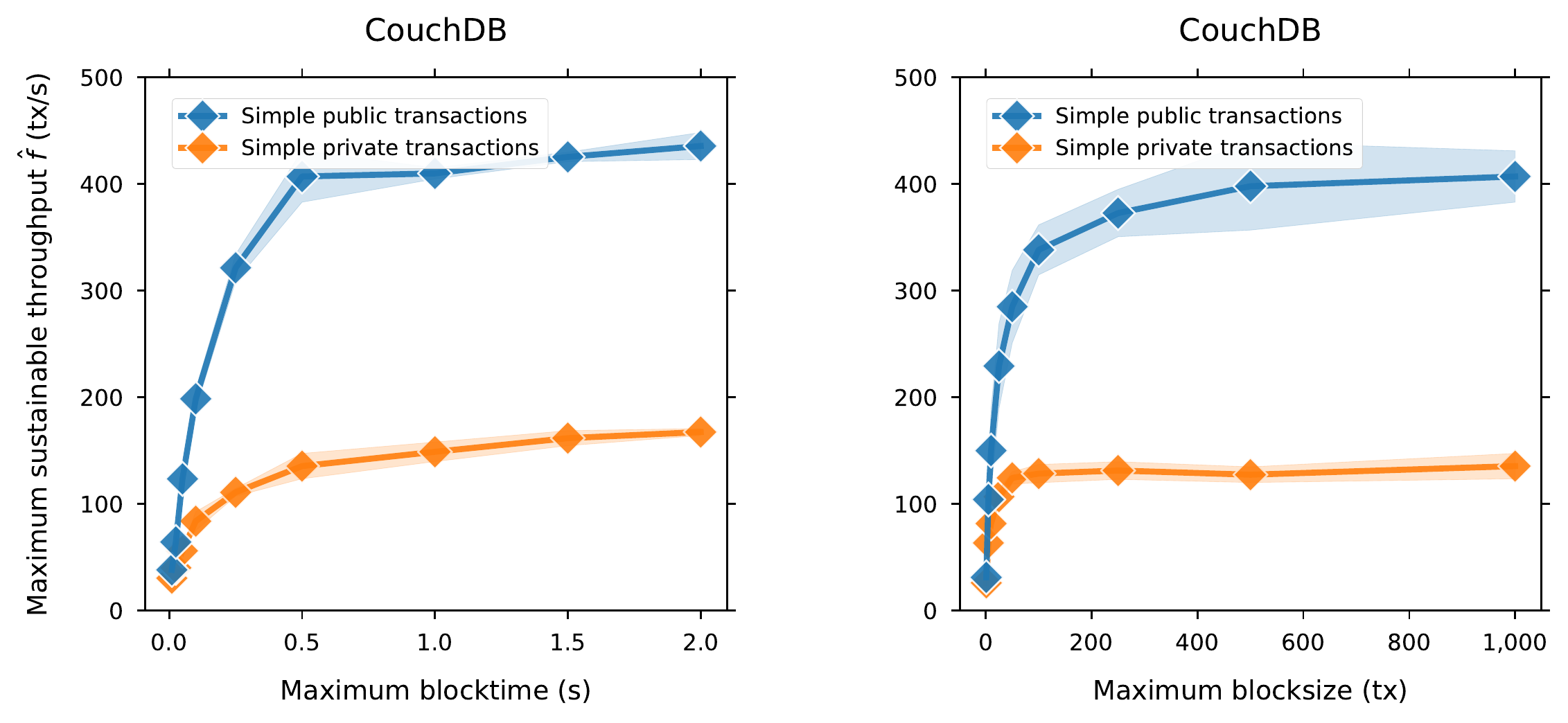}
    \includegraphics[width=0.8\linewidth, trim= 0cm 0cm 0cm 0cm ]{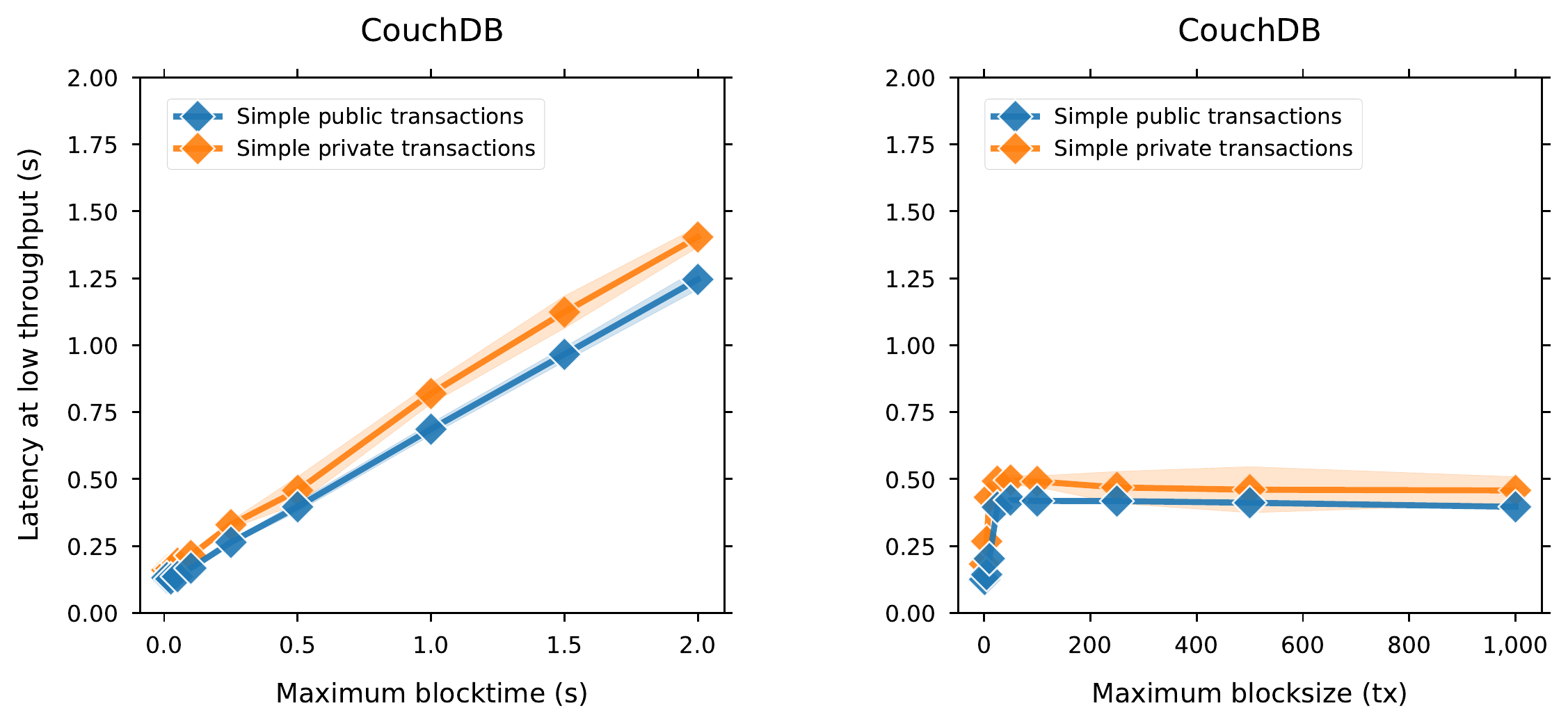}
    \caption{Comparison of different block times and block sizes.}
    \label{fig:def_batch_timeout}
\end{figure}

When we varied maximum blocksize, we got similar results, but with a ``cutoff''. This is because we used the default maximum blocktime of 0.5\,s, which -- considering that maximum throughput is around 500\,tx/s when blocks become sufficiently large -- becomes the actual trigger as soon as the maximum blocksize is higher than 0.5\,s$\,\cdot\,$500\,tx/s$\,=\,$250\,tx. For the low throughput tests on latency, i.e., at 50\,tx/s for public transactions and a blocktime of 500\,ms, blocks never get bigger than 25\,tx. Accordingly, we see no changes in latency beyond 50\,ms.
See Figure~\ref{fig:def_batch_timeout} for an overview of these results. 

\subsection{Business Logic}
\label{subsec:business_logic}

\subsubsection{I/O-Heavy Workload}

\begin{figure}[!tb]
    \centering
    \includegraphics[width=0.8\linewidth, trim= 0cm 0cm 0cm 0cm ]{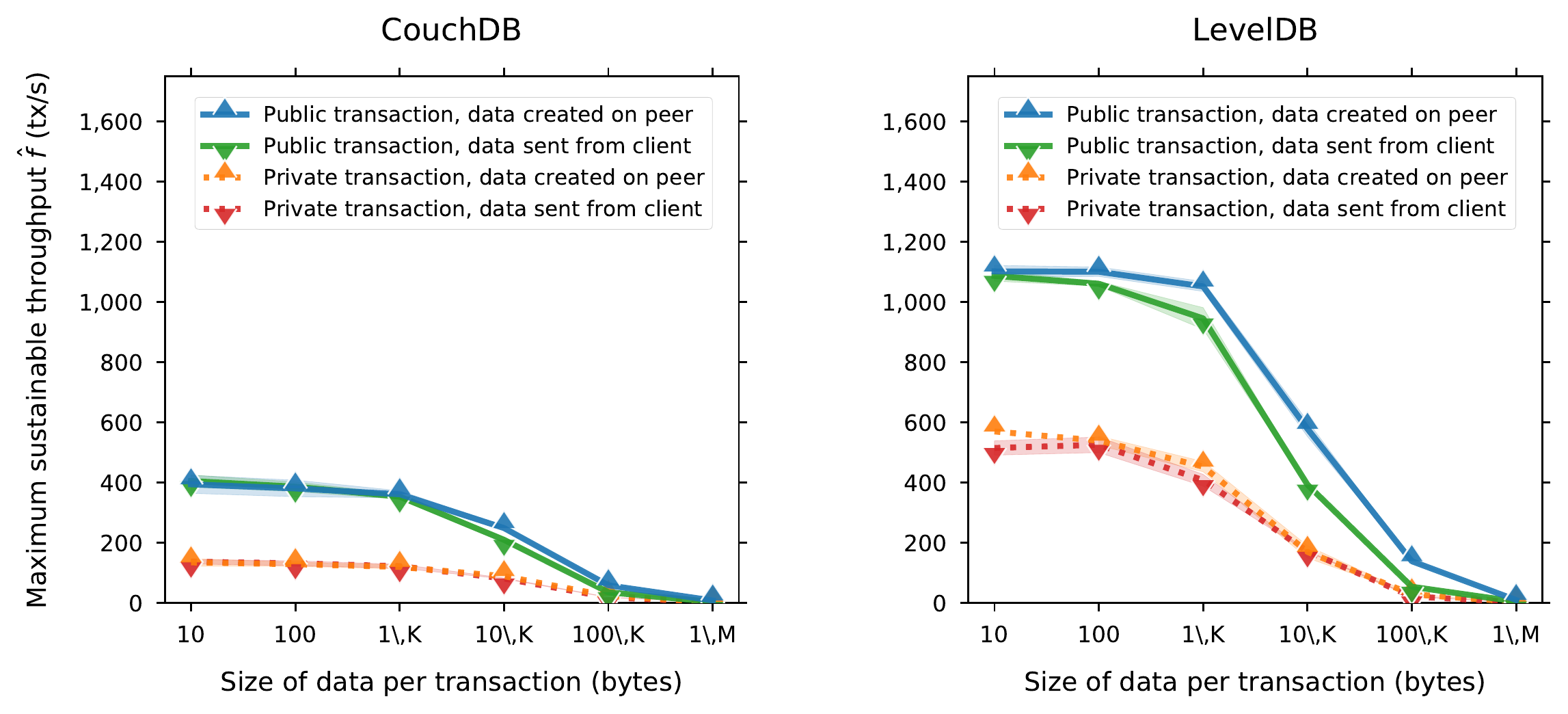}
    \caption{Comparison of performance with varying transaction size.}
    \label{fig:instance_type_latency}
\end{figure}

The first test we ran was on the impact of maintaining larger data sets in terms of the keyspace size of the state databases. We did not observe any relevant dependence on the keyspace size for less than $10^5$ keys (see Figure~\ref{fig:instance_type_latency}). Performance implications of very large keyspace sizes for LevelDB are given by the likes of~\citet{Blockbench:2017:Dinh,baliga2018performance}. Due to space restrictions, we consider this to be a property of the databases themselves rather than the \ac{hlf} network. 

Next, we checked how sensitive throughput reacts to changes in size of the data written in a single transaction, both when the data is communicated via the client (data sent from the peer) and when it is already present on the peer (for instance, created there a result of executing a smart contract). We observed that, as long as the bandwidth is adequate, like in a cloud data-center, it is not significant whether a large amount of the data that is to be processed is created on the peer or sent via the client. Transactions with 10~bytes have around the same throughput as the simple (public/private) transactions that have already been benchmarked before. Switching from 10~bytes to 1\,kB only causes degradations of less than 10\,\% for CouchDB and less than 20\,\% for LevelDB. However, moving from 10~bytes to 100\,kB degrades throughput by more than 85\,\% for public transactions (even 95\,\% if the data is generated on the client; which indicates that networking is also resource-intensive) and 75\,\% to 95\,\% for private transactions. We also noted that the degradation of CouchDB and LevelDB is similar, except for private transactions with CouchDB. Here, throughput is already rather low for 10\,kB. So, while there is no significant difference between the creation of the data on the client (networking intensive) and peer (no additional networking) for 10~bytes, it is 200\,\% higher for 100\,kB LevelDB public, private and CouchDB public. For private transactions with CouchDB, the difference is only 30\,\% -- 50\,\%. For 1\,MB, throughput is less than 10\,tx/s for LevelDB and less than 3\,tx/s for CouchDB. The maximum throughput in terms of data is approximately 14\,MB/s for the run with 100\,kB packages. This low sensitivity of throughput to transaction size up to the order of few kilobytes may be due to the certificate handling of \ac{hlf}: Each transaction carries the digital certificates of the client that submits it and the peers that endorse it. We measured that these certificates -- including the sub-certificate of the corresponding organization and the root certificate of the certificate authority -- have a size of approximately 1\,kB. Accordingly, as long as the payload is smaller than~1\,kB, it has a negligible impact on peers' and orderers' networking effort.

\subsubsection{Reading Data}

First, we checked that the keyspace size does not impact fewer than $10^5$ keys. Reading speed is only a reasonable number on a ``per peer'' basis because no other node is involved in a reading operation (except for cross-checks if the client does not trust the peer). For simple key-based queries on m5.large instances, we obtained approximately 400~reads per second on CouchDB (150~reads per second with complex queries) and around 750~reads per second on LevelDB. We used non-invoked queries that do not involve the \ac{hlf} transaction flow from Section~\ref{sec:background}. Again, we used the standard configuration, consisting of four clients and two peers, in accordance with which clients distribute requests equally between the peers.

Complex queries are only feasible on CouchDB. Here, we could observe a massive difference between no indexing (which performs approximately as badly as querying the total database and searching the value space afterward, resulting in a low one-digit number of successful queries per peer and second) and indexing, which still allows approximately 150~reads per second and peer. It is worth noting that networks with high-performance requirements on reading processes should either opt for multiple peers for scaling benefits or consider retrieving the peers' data and maintaining a separate database for queries.

\subsubsection{CPU-heavy workload}

\begin{figure}[!tb]
    \centering
    \includegraphics[width=0.8\linewidth, trim= 0cm 0cm 0cm 0cm ]{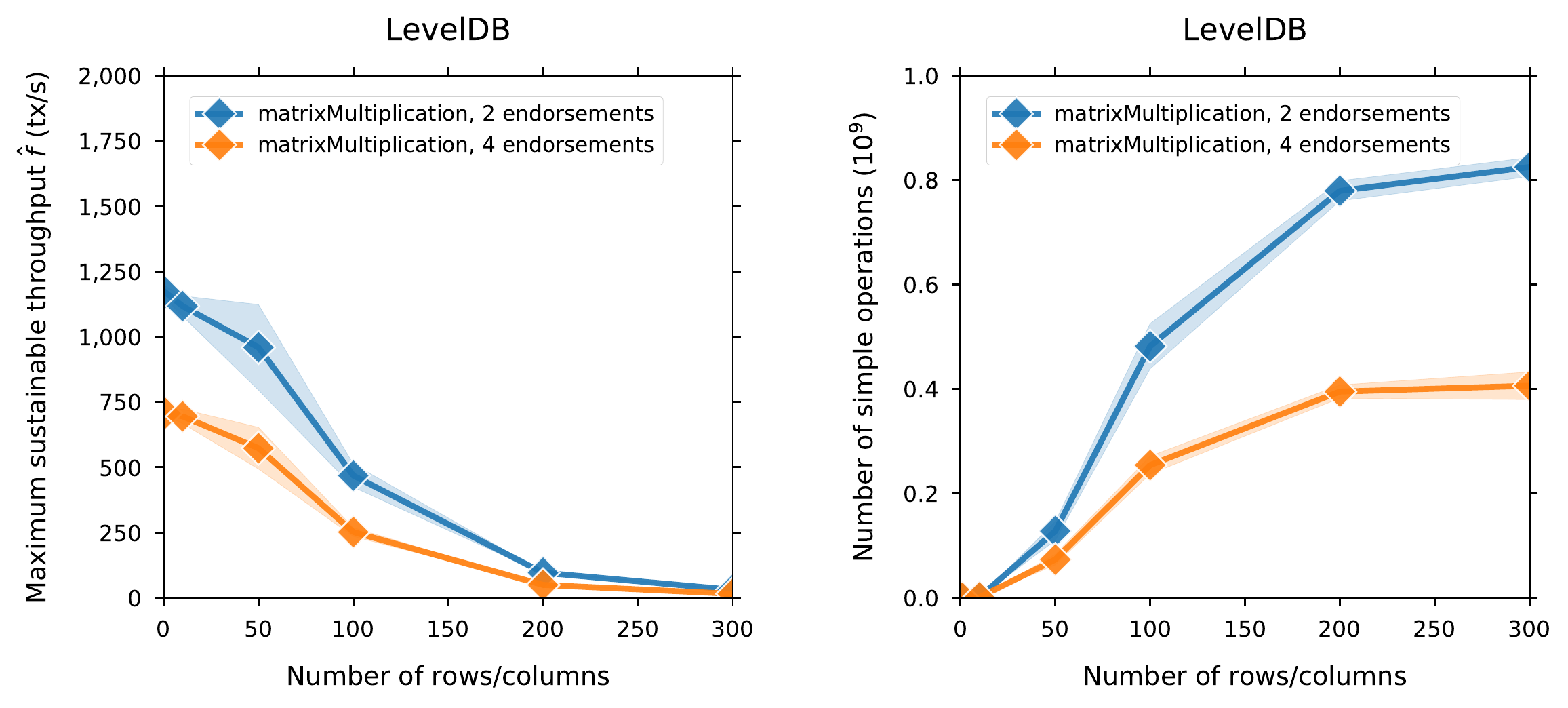}
    \caption{Different tasks difficulties by matrix multiplication.} 
    \label{fig:throughput_matrix_multiplication}
\end{figure}

To test \ac{hlf}'s performance on CPU heavy operations, we conducted matrix multiplications which we implemented through simple nested loops, with different matrix sizes as this allows for quantitative control of complexity. Please see Figure~\ref{fig:throughput_matrix_multiplication} for an overview of our findings.
Multiplying two n$\times n$ matrices requires $\mathcal O$(n$^3$) simple operations (additions and multiplications) in our nested loop implementation. So, for large n, we expected the throughput to scale as $\tfrac{1}{n^3}$. Indeed, we see that the total number of operations approached a saturation curve for large n. In contrast, we found that, for small n, the \ac{hlf}-related overhead matters.
For n=300, the performance of the network is still around 30\,tx/s respectively 15\,tx/s for two respectively four endorsements. As we could measure, this corresponds to the performance of a standalone node.js application that runs the nested loops. To be more precise, in \ac{hlf}, the chaincode is run in separate docker containers that communicate with the peer container, so every endorsing peer's associated chaincode container executes the code. During validation, the peers then check that the simulation results of the endorsing peers coincide. This means that, leaving aside the degree of redundancy determined by the endorsement policy, the chaincode (smart contract) can run code as efficiently as a centralized system when the blockchain-related operations (networking, signatures) are negligible in comparison to the overhead caused by networking and committing operations in \ac{hlf}. In contrast, a matrix multiplication that is also implemented through nested loops and run in the Ethereum Virtual Machine cannot deal with a multiplication of a 90$\times$90 matrix. Furthermore, multiplying a 30$\times$30 matrix takes almost an entire second. With regard to executing CPU-intensive tasks, then, all of these data points illustrate the significant performance benefits of using Fabric compared to (both private and public) Ethereum-based blockchains when executing CPU-intensive tasks.

As expected, we were able to confirm that there is no difference between public and private transactions for a matrix multiplication since there are no database operations involved. Moreover, for a stricter endorsement policy (four out of eight), throughput is approximately half of that attained with a weaker endorsement policy (two out of eight) because, in total, there are twice as many computations for a single transaction. When comparing four endorsements and two endorsements, the ratio of maximum throughput is 40\,\% for multiplying a 1$\times$1 matrix, 46\,\% for a 100$\times$100 matrix, and 50\,\% for a 300$\times$300 matrix. With the initial presence of \ac{hlf}-related overhead for small matrices, we hence get the expected asymptotic value. 

\subsection{Network}
\label{subsec:network}

\subsubsection{Delays}
\begin{figure}[!tb]
    \centering
    \includegraphics[page=4, width=0.85\linewidth, trim=1cm 3cm 1cm 3cm, clip]{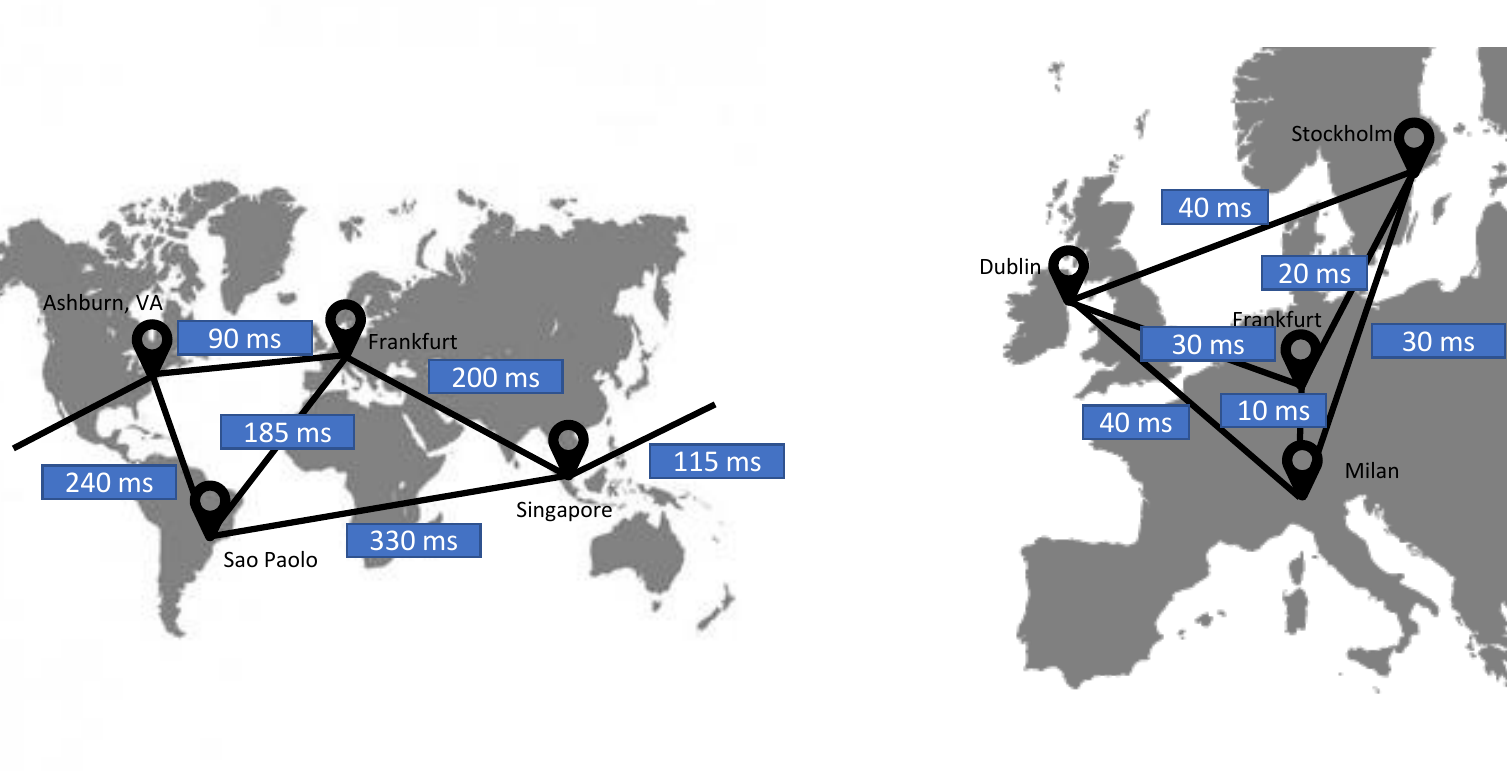}
    \caption{Network topologies and corresponding network delays (one-way) used to determine the impact of network topology on maximum throughput.}
    \label{fig:topology}
\end{figure}

To investigate the impact of network delays in a sufficiently general real-world scenario, we defined groups within our default architecture, where each group corresponds to an organization, representing an enterprise and consisting of two peers, one orderer, and four clients. We proceeded on the assumption of minimal network delays within a group. While this hypothesis is certainly optimistic for global enterprises, if speed is of the essence in a large network, one may well choose the nearest peers within an organization for endorsement. In a first attempt, we used the standard traffic-control (tc) tool available on Ubuntu servers to set an artificial delay for any communication between the members of different groups. However, we noticed that the results obtained by imposing artificial delays became very unreliable at high throughput, which indicates that, when CPU utilization or network traffic is high, tc~does not operate reliably. To address this, we started using deployments over multiple data-centers and set up a cross-European and global network. Specifically, we set up groups in Germany, Ireland, Italy, and Sweden for the European case with moderate network delays, and in Germany, Brazil, Singapore, and the East of the US for the intercontinental case with high network delays. We noted that latency increases by 30\,\% to 50\,\% from a single datacenter to a cross-European (30\,ms one-way network delay) and by more than 200\,\% from a single data center to an intercontinental distributed system (up to 330\,ms one-way network delay). In the intercontinental case, transactions will, on average, take 1.2~seconds (public) and more than 1.7~seconds (private), even at low throughput. Once throughput approaches maximum sustainable throughput, the latencies become even higher. A detailed topology of the network, including the network delays we measured between each data center pair, can be found in Figure~\ref{fig:topology}.

\begin{figure}[!tb]
    \centering
    \includegraphics[width=0.8\linewidth]{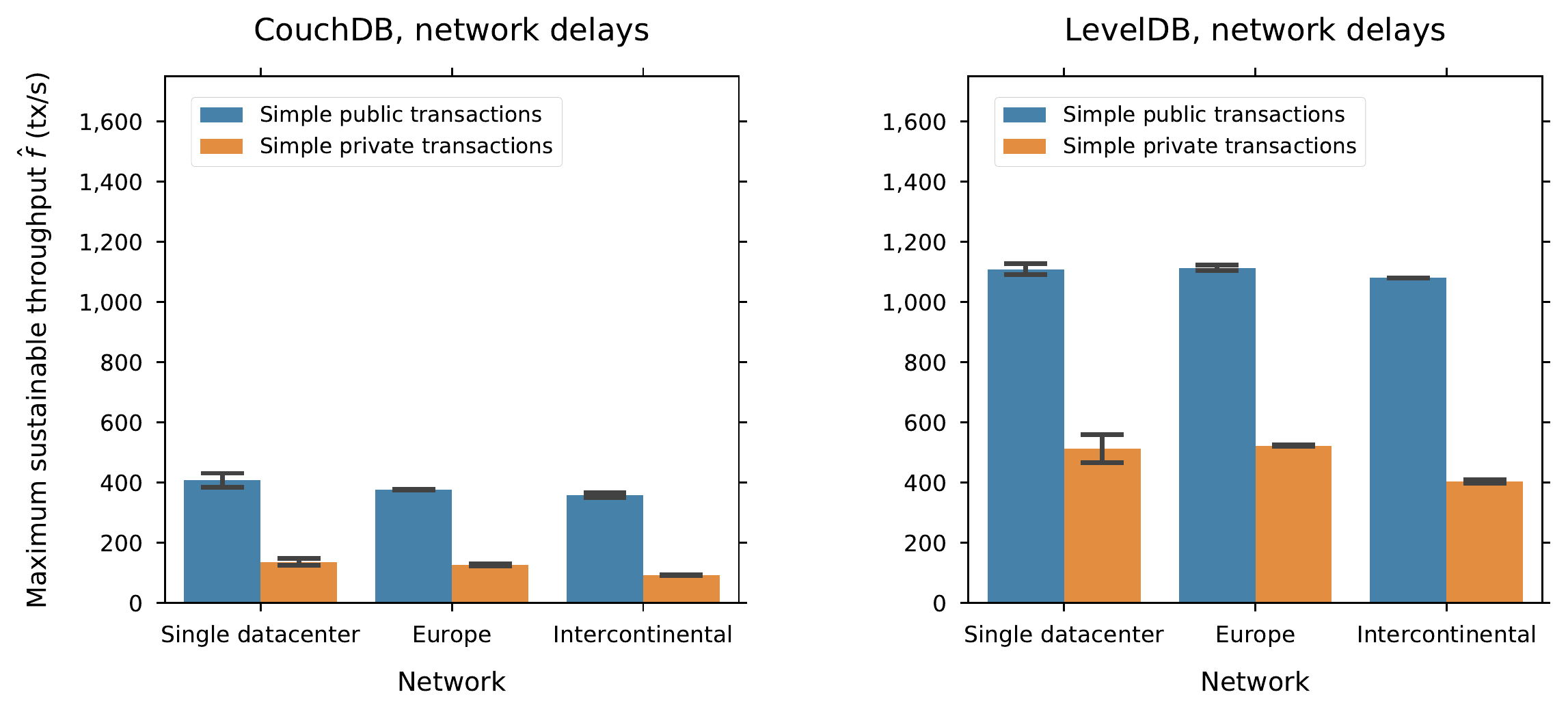}
    \caption{Maximum throughput for different geographical distributions of nodes.}
    \label{fig:intercontinental}
\end{figure}

Having imposed artificial network delays by using tc for our initial simulation, we noted a decrease in performance by approximately 50\,\% for CouchDB and 70\,\% for LevelDB as well as a significant standard deviation thereof for delays of 50\,ms. Meanwhile, by using the actual cross-data center deployments with real-world delays, we found that, for both LevelDB and CouchDB, and indeed for both public and private transactions, performance does not degrade as significantly in the intercontinental case (see Figure~\ref{fig:intercontinental}). This lends further credence to a statement by~\citet{androulaki2018hyperledger}, according to which a cross-data center deployment of a large number of nodes still offers high performance. In their experiments, the authors deployed 100~nodes, located in five different datacenters, and used LevelDB for the peers’ databases. In our own experiments, we found that, for public transactions with CouchDB, maximum throughput decreases from 426\,tx/s for the single datacenter case to 376\,tx/s in the cross-European case and 358\,tx/s in the intercontinental case. This corresponds to a drop of 12\,\% and 16\,\%, respectively. We also observed that the performance decrease is less significant for LevelDB in the cross-European case. In contrast, for both CouchDB and LevelDB, the throughput decrease of private transactions in the intercontinental case is considerable. Indeed, for CouchDB, we observed a drop of 39\,\% in maximum throughput compared to the single datacenter case. For LevelDB, we noted a drop of 26\,\%. We attribute this increased latency sensitivity of private transactions to the additional networking required in order to distribute the payload to the eligible peers (see Section~\ref{subsec:private_flow}).

To conduct a systematic investigation of the relationship between performance metrics and network delays, we had to adapt our benchmarking procedure. The same was required for a latency analysis, which we found to be (intuitively and empirically) far more sensitive concerning network delays than throughput. While the real-world deployments make it difficult to vary network delays continuously, we found that the latencies in those real-world deployments are similar to the latencies we observed when we stayed well below maximum throughput in our measurements. 
By conducting corresponding experiments with artificial network delays imposed by means of the tc tool, we found that transaction latency increases proportionally with network delays;. Interestingly, the average slope in this case is approximately~15. In other words, an increase in latency $\Delta $l for bilateral communication between servers (including clients, peers, and orderers) that belong to different organizations implies an increase of around $15\times\Delta l$ for the latency of transaction confirmation, i.e., from triggering the transaction request on the client until receiving a confirmation on the client that the transaction has been committed in a peer’s ledger. This suggests that there are approximately 15~communications between different nodes in the lifecycle of a single transaction. Since \ac{hlf} networks have to meet high performance requirements, it is especially important to avoid communication paths with notable network delays. This can be achieved, for instance, by weakening endorsement policies and choosing endorsers with low network latency, or by avoiding particularly large distances between ordering nodes. Opting for close proximity between nodes, however, can negatively affect their availability guarantees (``liveness'') because stronger geographic localization increases the threat of correlated crash-failures as, for instance, caused by blackouts in power grids.

\begin{figure}[!tb]
    \centering    
    \includegraphics[width=0.8\linewidth, trim= 0cm 0cm 0cm 0cm]{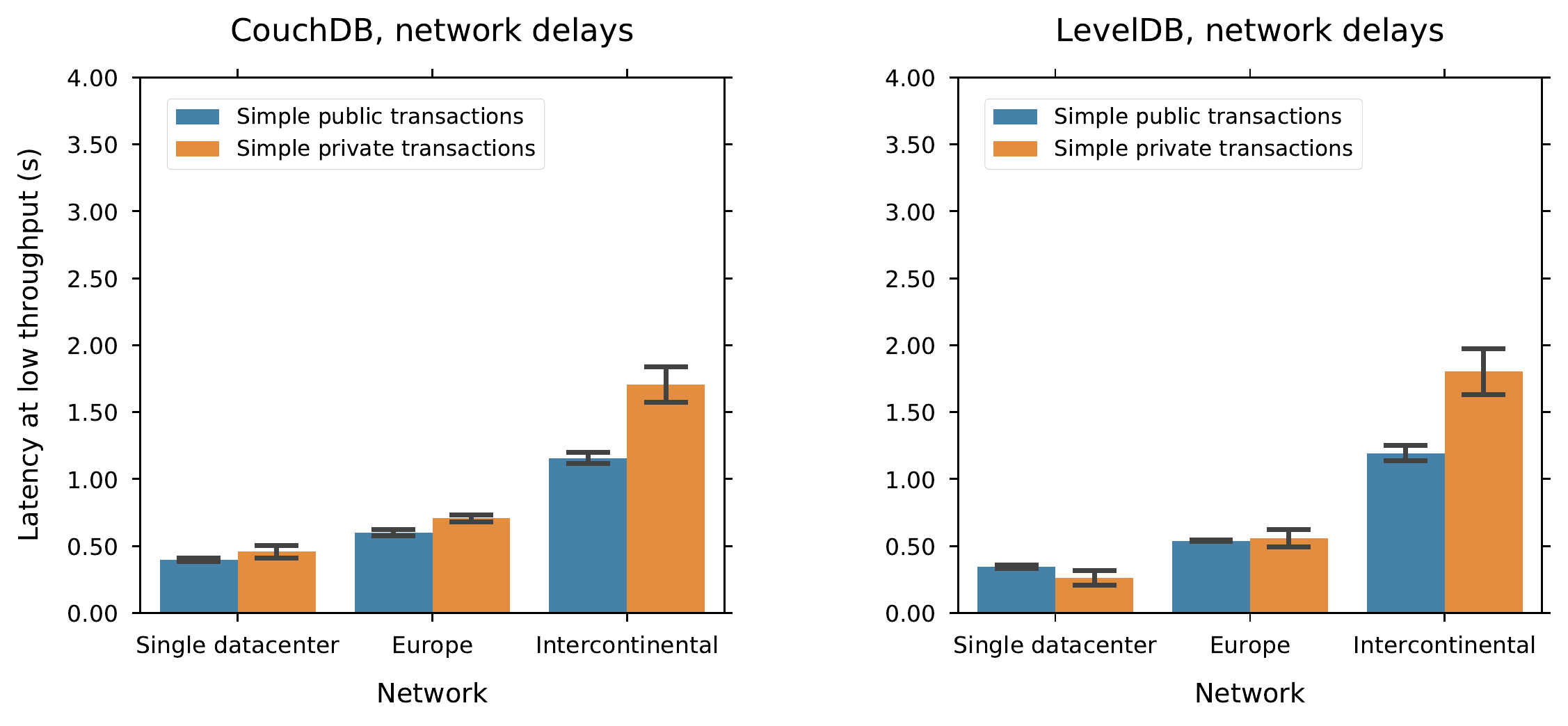}
    \caption{Latency for different geographical distributions of nodes.}
    \label{fig:latency_delay}
\end{figure}

\subsubsection{Bandwidth}
\label{subsec:bandwidth}

\begin{figure}[!tb]
    \centering
    \includegraphics[width=0.8\linewidth, trim= 0cm 0cm 0cm 0cm ]{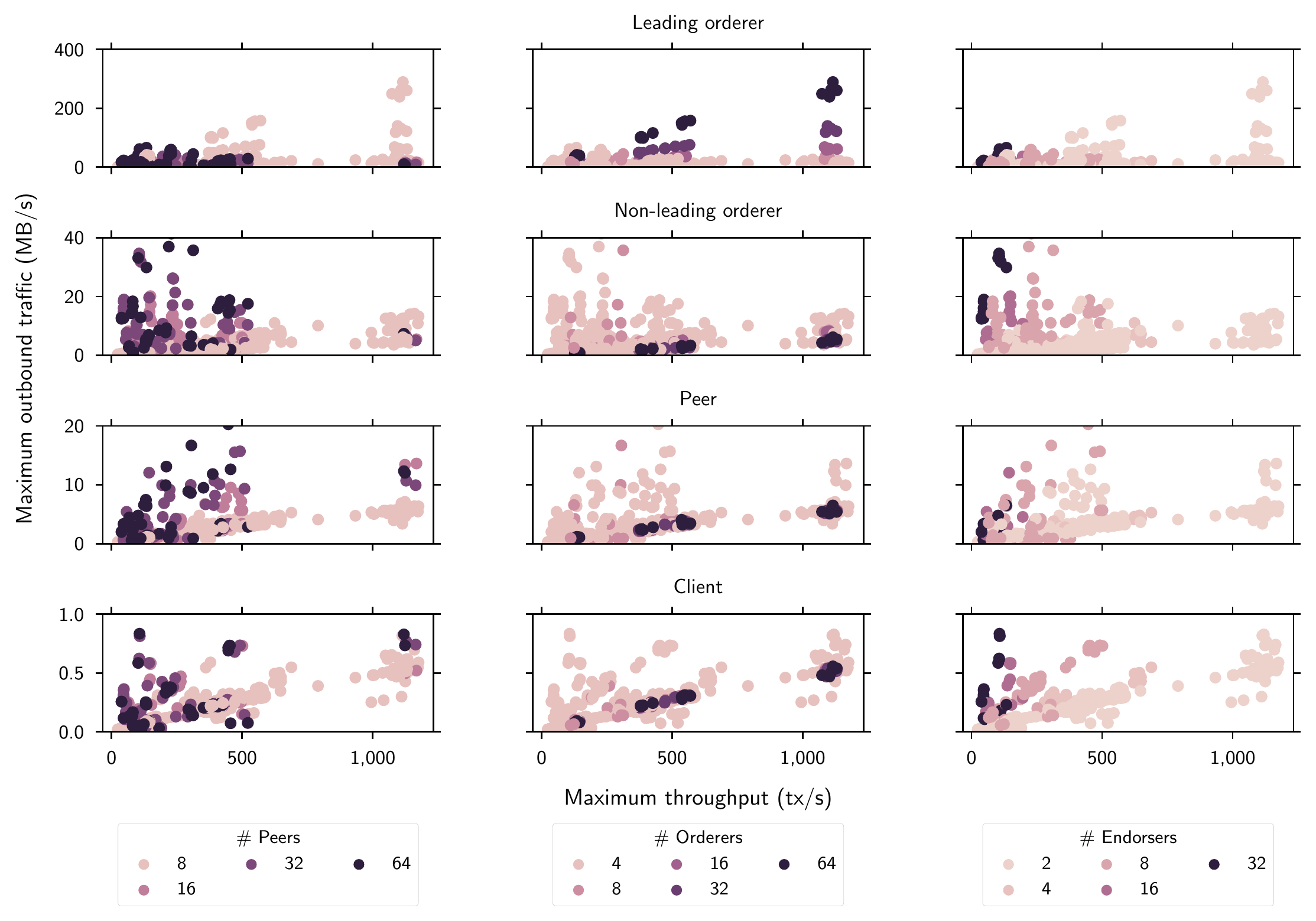}
    \caption{Required bandwidth for different roles and architectures.}
    \label{fig:bandwidth}
\end{figure}

Within a \ac{hlf} network, each role requires certain bandwidth, so we investigated those requirements with regard to orderers, peers, and clients. For peers and clients, we found that inbound traffic is distributed uniformly. Moreover, the maximum requirement on download speed is homogeneous for peers and clients, as it is for orderers. The maximum values we observed were the same as the respective maximum on outbound traffic, and since upload speed is more likely to create a bottleneck than download speed, we will focus our discussion on the requirements concerning upload. According to intuition, as Figure~\ref{fig:bandwidth} illustrates with regard to these three roles in the network and different architectures, there is a general linear correlation between throughput and outbound traffic for all roles. In their \ac{hlf} network, \citet{Thakkar:2018:Fabric} measured the download rate of a peer to be approximately 2.5\,MB/s (and the download rate 0.5\,MB/s). Regarding the upload rate of peers, we arrive at a similar order of magnitude for equally high throughput. 

In contrast, we found the upload rate of orderers to be more heterogeneous, and at times significantly more sizeable. To be specific, the RAFT leader requires a very high upload speed when the ordering service has multiple nodes. For n$=$64 orderers, for example, we observed an upload rate of more than 350\,MB/s (bearing in mind that maximum throughput is independent of the number of orderers for up to 64 orderers, upload is still not the bottleneck, at least not for deployment in a single datacenter with high networking capabilities). This is plausible because the crash-fault tolerant consensus mechanism RAFT, which Fabric uses for the ordering service, follows a two-phase commit paradigm. Therefore, the complexity of network traffic, i.e., the number of sent messages, is in the order of n$\times$(n$-$1), and the leader needs to be involved in all of these messages. For the other orderers, outbound traffic is one order of magnitude smaller. As the charts in the second row of Figure~\ref{fig:bandwidth} indicate, the upload speed of non-leading ordering nodes depends mainly on the number of peers in that network, as well as on the number of endorsers per transaction, which both makes sense as they need to distribute new blocks to the peers. After all, transactions are larger when more endorsements (signatures) have to be collected.

As rows three and four row of Figure~\ref{fig:bandwidth} indicate, this observation also holds true for the upload requirements of peers and clients. Moreover, for the clients, the linear interrelation between outbound traffic and maximum throughput is clear. The upload speed requirements on non-leading orderers are often approximately twice the requirement on the peers. This makes sense as there were twice as many peers as orderers in our default scenario. It is further worth noting that the clients only have a very small requirement concerning outbound network speed. 

\subsection{Robustness}
\label{subsec:robustness}

\subsubsection{Temporal Distribution of Requests}

As illustrated in Section~\ref{sec:methods}, the \ac{dlps} sends transaction requests highly uniformly by default. When we tested different temporal distributions of the requests (i.e., jitter), we modified this default to a step-shaped distribution in order to evaluate the queuing system's sensitivity and efficiency. Here, clients send transactions at the beginning of each second and they do so with a fluctuating distribution that has notably more or fewer transactions per second ($\Delta\leq\tfrac{f}{2}$). In this scenario, we did not notice a significant deterioration of maximum throughput and latency. This suggests that, as long as queues do not become too long, the queuing process of \ac{hlf} is efficient.

\subsubsection{Node Crashes}

\begin{figure}[!tb]
    \centering
    \includegraphics[width=0.8\linewidth, trim= 0cm 0cm 0cm 0cm ]{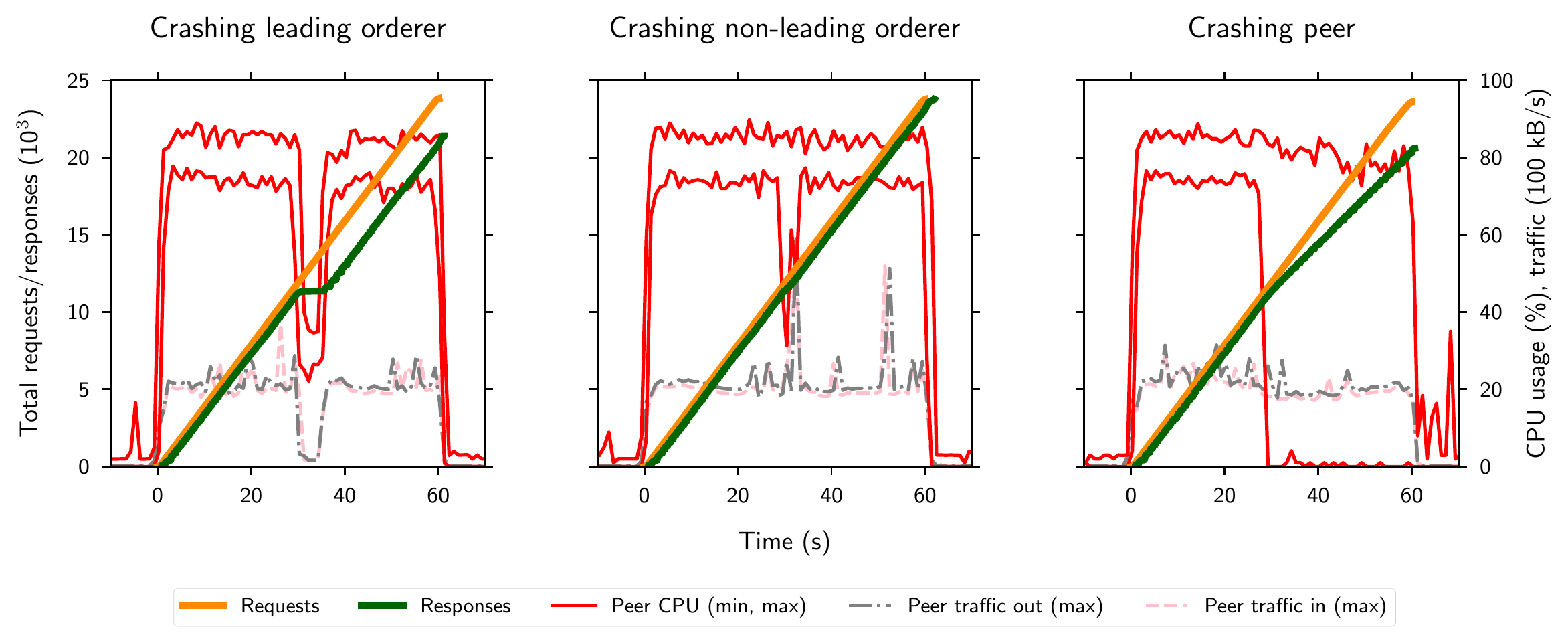}
    \caption{Performance impact of crashing leading or non-leading orderer nodes and peers (at t$=$30\,s) on performance.}
    \label{fig:crashes}
\end{figure}

As soon as a system transitions from testing to productive usage, its resistance and resilience against failures become matters of great relevance. By operating multiple peers within one organization on physically separated nodes with the use of a blockchain, the risk of data loss caused by crashes and attacks is already notably mitigated. Similarly important, however, is how the failures of individual nodes affect overall performance, since it might take some time until a failed node is compensated or reset and re-synchronized. For the purpose of this study, we examined the different roles in the system with the expectation that each would have different failure ramifications. We focused on crashes because malicious attacks require sophisticated and specialized implementations and -- since we are in a private permissioned network -- can be traced to the responsible parties and therefore deterred. While this will no doubt make an interesting topic for future research, we will not look into it any further here, nor into crashes of clients since we used enough clients to saturate the system and clients can easily be replaced on short notice as there is no need for synchronization. Therefore, the relevant aspects for this study are the crashes of orderers and peers. Since the recommended ordering service is the crash fault-tolerant RAFT, we expected that crashing a single orderer would not significantly impact a \ac{hlf} network's operation. Figure~\ref{fig:crashes} depicts the impact of crashing various node types.

To test the outcomes, we put the network under stress at 400\,tx/s, which is close to maximum throughput (and therefore maximum CPU utilization). After 30~seconds of sending transactions, we crashed a single orderer and continued sending requests at the same rate for an additional 30~seconds. We found that the total impact of crashing an orderer is indeed limited. However, it makes a considerable difference whether the crash affects the current RAFT leader or a non-leading orderer: When the crash affects a non-leading ordering node, the ordering service stops distributing new blocks for around 5\,seconds, whereupon it resumes at the previous speed with a newly selected orderer (Figure~\ref{fig:crashes}, chart on the left). If a non-leading orderer crashes, however, the impact on performance is negligible (Figure~\ref{fig:crashes}, chart in the center). If a single peer crashes, the performance drops by the rate of transactions that required the respective peer as an endorser. However, this is predicated on the fact that we limited clients to requesting endorsements from a fixed set of peers that contained exactly as many peers as required by the endorsement policy. In this case, we used our default configuration with four organizations, each associated with two peers, and two endorsements for every transaction. Consequently, every peer participates in $\tfrac14$ of all transactions, which explains the drop of throughput by 25\,\% after 30~seconds. In a production-grade \ac{hlf} network, one would likely provide each client with at least a few more peers to compensate for crashes, which would prevent failing transactions. However, shifting the endorsement workload to another peer may decrease maximum throughput to that of a \ac{hlf} network without the crashed peer. For a comprehensive discussion of other reasons for error-prone transactions, see~\cite{chacko2021fail}.

\section{Discussion}
\label{sec:discussion}

\ac{hlf} is a highly customizable permissioned blockchain framework that allows enterprises to adjust the network architecture to their requirements. While this ability allows for many optimizations, it also leads to complexity and requires in-depth knowledge of \ac{hlf}’s design options and parameters as well as their performance impact. The purpose of this paper is to link in with current research in efforts to advance the understanding of the key metrics one has to consider when setting up a \ac{hlf}-based application. Having managed to reproduce many of the results of these related works, we broadened our expertise in \ac{hlf} with a bilateral strategy; by using a benchmarking framework built on precise definitions of key metrics and by testing many as yet unexplored settings in a structured manner. For instance, we worked with the findings of \citet{androulaki2018hyperledger} concerning the effect of network delays on the throughput of \ac{hlf}, but we extended their results by comparing three different setups: no delay, continental, and inter-continental. This allowed us to understand how \ac{dlps} can be used to run tests for a wide range of variables in order to evaluate a blockchain-based system’s performance potential prior to implementation. Please see Table~\ref{tab:results} for details on each contributing factor.

\begin{table}[!htb]
\centering
\resizebox{0.85\columnwidth}{!}{
\begin{tabular}{L{2.5cm}|L{4cm}|p{11.5cm}}
\midrule
\bf{Group} & \bf{Impacting Factor} & \bf{Results} \\\midrule
\multirow{4}{*}{Architecture}
& Number of organizations, peers, and orderers & In our experiments, the number of orderers does not affect overall performance. Adding peers to small networks while keeping the endorsement policy constant improves performance. When a large number of organizations gets involved, performance degrades, albeit to a rather small extent at first but increasingly so with bigger networks.\\
& Endorsement policy & A higher number of required endorsers reduces total throughput. Introducing additional peers can mitigate this.\\
& Number of channels & The number of channels has little impact on the performance of the system.\\
& Database location &  Moving the database to another server offers limited benefits for throughput.\\
\midrule
\multirow{3}{*}{Setup}
& Hardware & Increasing the number of vCPUs increases throughput significantly for fewer than eight vCPUs but less for a larger number of vCPUs.\\
& Database type & The database type has a significant impact on system performance. With LevelDB, throughput is up to three times higher than with CouchDB.\\
& Block parameters & Block time of around 0.5 s yields a particular sweet spot. Any addition of block time or block size, respectively, has only limited performance benefits but increases block latency.  Below a block time of 0.5 s, throughput
decreases considerably.\\
\midrule
\multirow{4}{*}{Business Logic}
& Private data & Throughput for public transactions is around three times higher for CouchDB and around two times higher for LevelDB than for private transactions. \\
& I/O-heavy workload & Once the transaction payload is larger than~1\,kB, throughput decreases rapidly. Total upload is bounded to tens of MB/s.
\\
& CPU-heavy workload & CPU-heavy node.js smart contracts work as fast as native implementations. \\
& Reading vs. writing & Read throughput scales linearly with the number of peers while write throughput depends on many parameters.\\
\midrule
\multirow{2}{*}{Network}
& Delays & The impact of network delays on throughput and latency is relatively low, even in an intercontinental network. This impact is greater on private data than it is on public data.\\
& Bandwidth & The bandwidth requirements rise proportionally to the number of nodes. Considering the RAFT setup, the leader node demands a comparatively higher upload bandwidth with an increasing number of orderers.\\ 
\midrule
\multirow{2}{*}{Robustness}
& Node crashes & \ac{hlf} is very robust with regard to crashes. A crashing peer does not affect the total network beyond its loss in endorsement power. If a Raft leader crashes, it takes about 5s for the system to resume normal operations.\\
& Temporal distribution of requests & Small deviations in distribution do not impact system performance as long as peaks stay below the maximum sustainable throughput.
\end{tabular}
}
\vspace{5pt}
\caption{Results of our benchmarking study by impacting factor.}
\label{tab:results}
\end{table}

\begin{figure}[tb]
    \centering
    \includegraphics[width=0.85\linewidth, trim= 0cm 0cm 0cm 0cm ]{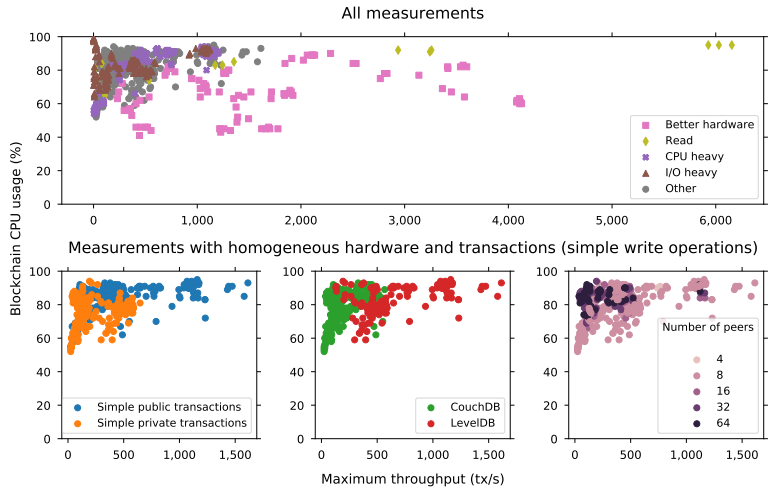}
    \caption{Summary of our measurements and the most important design parameters.}
    \label{fig:all_throughput_cpu}
\end{figure}

As the overview of our measurement results in Figure~\ref{fig:all_throughput_cpu} shows, maximum throughput depends largely on the type of transaction (reading operations, CPU heavy transactions, i/o heavy transactions, and simple write transactions) and the type of hardware. For homogeneous hardware (such as m5.large, for which we conducted the most experiments), there is a clear correlation between maximum throughput and CPU use across highly heterogeneous deployments. We found that both are heavily dependent on the kind of database used (LevelDB achieves higher throughput), the visibility of transactions (private transactions achieve lower throughput), and network size (large \ac{hlf} networks have lower throughput). Therefore, these parameters should be considered with particular attention to detail when conceptualizing the network architecture for a use case with high performance requirements.

\citet{kannengiesser2020} describe various trade-offs that developers have to make when working on blockchain systems. This study examines some of these trade-offs and provides additional metrics to quantify them in the case of \ac{hlf}. In particular, our measurements of different Smart Contract methods, such as varying the complexity of matrix multiplications and the size of transactions, quantify the trade-off between transaction validation speed and operation complexity.
Similarly, by investigating private and public transactions in \ac{hlf}, we also quantify the trade-off between confidentiality and performance suggested in article of ~\citet{kannengiesser2020}. Finally, our various performance measurements of varying network sizes, topologies, and endorsement policies quantify the extent to which performance depends on decentralization and with it security and availability. As our experiments demonstrate, the ordering service is not the bottleneck in the examined architecture, so the trade-off between performance and security was only present in the endorsement policy. The solo orderer, lacking any crash or Byzantine fault tolerance, provided approximately the same overall performance as the crash fault-tolerant RAFT ordering service. It will be interesting to see whether the future use of a Byzantine fault-tolerant ordering service will have any impact.

Our results have several implications concerning the use of blockchain for supply chain management. First, we were able to validate the theoretical performance of \ac{hlf} in that it supports up to several hundred or indeed a few thousand transactions per second. Nevertheless, due to the current implementation of gossip dissemination, we found that the performance decreases when \ac{hlf} serves a large number of organizations. Private blockchain systems based on \ac{hlf} should, therefore, avoid the integration of too many organizations. Second, while private data is an important function of supply chain information sharing systems~\citep{guggenberger2020improving}, we found that it also reduces overall performance to a surprisingly large extent. With this in mind, we propose to use this function only where necessary. At all other times, standard public data transactions are to be given preference, or \ac{hlf}-based applications ought to be supplemented with bilateral communication via standardized APIs to avoid overloading the blockchain. Third, the same applies to the database type. Advanced blockchain provenance solutions can use CouchDB’s query capabilities to quickly extract data from the system. While CouchDB offers fast queries, its use considerably degrades system performance. Therefore, its use is only advisable when absolutely necessary. 
 
Another solution, and one that a future iteration of \ac{hlf} could turn into a desirable feature, is that an organization can also keep the world state in another database by periodically pulling an update from the database  natively supported by \ac{hlf} peers. This other database could be optimized for the queries required by the use case, for example, a graph database for supply chain use cases that record the successive joining of components. This approach would also make it possible to decouple query operations from the peer tasks. Doing so would ensure that no queries could impact system performance. Since large network sizes degrade throughput, another useful approach would be a horizontal scaling of databases associated with one peer, instead of deploying several peers, but this would only be advisable if an application is expected to have a substantial query throughput.

Finally, despite the limitations that became apparent in our benchmarking results, \ac{hlf} performs exceedingly well on tasks that require extensive computation in an inter-continental environment. The impact of network delays is limited, which puts a global infrastructure, as proposed by TradeLens \citep{jensen2019tradelens}, within the realm of feasibility.

Here, we have focused on a subset of interesting factor combinations because the large number of parameters that can be configured in a \ac{hlf} network makes it impractical to test all possibilities. We settled on a standard configuration and changed individual parameters to identify their impact on performance. It is important to note, therefore, that fellow researchers who depart significantly from our testing scenario might attain somewhat different results. To name but one example of this potential deviation,~\citet{thakkar2021scaling} suggest that certain characteristics might change for very strong servers.

Accordingly, the results of this study are to be understood as indications of the potential of \ac{hlf}, rather than as strict reference points for all possible cases. We suggest that those who wish to move ahead to operational systems should first conduct a specific evaluation in which to include their respective particularities. When doing so, companies are welcome to use our extended version of the \ac{dlps} framework to test a wide range of variables and verify their blockchain project’s feasibility. Ultimately, the suitability of \ac{hlf} for business processes is case-specific. For example, we see an opportunity to use \ac{hlf} for the tracking of high-value goods, such as medical devices, in a B2B environment. However, other use cases may have requirements that go beyond \ac{hlf}’s capabilities. In particular, use cases with extreme spikes in transactions may be unsuitable for \ac{hlf}. On Singles Day in 2020, for instance, there were periods in which the online Marketplace Alibaba received almost 600,000 orders per second \citep{forbes2020}. In its current iteration, \ac{hlf} would not be able to process that many orders.

\section{Conclusion}
\label{sec:conclusion}

The purpose of this paper has been to examine the performance of \ac{hlf}. It provides an in-depth analysis of the system, covering a total of 15 variables concerning architecture, setup configuration, business logic, network, and robustness. The benefit of this analysis is the guidance it offers system architects and infrastructure engineers when designing \ac{hlf}-based infrastructure and applications.

To differentiate between the various theoretical and practical contributions of this study, it is important to note that, from an academic perspective, it contributes to the current understanding of how to design blockchain systems. It does this by exploring the full potential of private permissioned blockchains. We expect that future research will also benefit from the extended list of contributing factors by using it as a baseline for performance analyses of \ac{hlf} and other blockchains. As this study has indicated, in-depth performance analyses should incorporate corresponding measurements of the impact of multiple parameters. From a practitioner’s point of view, this study should be of value for demonstrating the impact of various parameters. This should help to optimize existing applications so as to facilitate higher network performance. Finally, by discussing the potential of \ac{hlf}, we provide a practice-oriented foundation on which practitioners should be better able to understand whether blockchain might meet their requirements for any potential applications at the level of their operations. They may also benefit from using our extension of the \ac{dlps} to benchmark their respective chaincodes and determine the feasibility of productive use in networks with specific parameters.

As our results demonstrate, \ac{hlf} is suitable to support the needs of many real-world industrial blockchain applications as it provides sufficient scalability along with multiple other key properties, such as ensuring high availability guarantees, accountability, as well as to some extent robustness to manipulation attacks. 
Owing to the large number of system parameters in \ac{hlf} that -- according to the measurements that we present in Section~\ref{sec:results} -- it is challenging to provide a universal and succinct quantitative assessment of \ac{hlf}'s performance; and a pre-study that considers the specific parameters in an enterprise application is advisable. Yet, we try to give a short summary: Even in large or intercontinental networks, \ac{hlf} can still reach more than 1000\,tx/s for public transactions with LevelDB. Using CouchDB and private transactions, each lowers throughput levels by a factor of three lower. The differences become smaller when database operations become less relevant, e.g., for large numbers of endorsers or complex workloads. This means that enterprise consortia can deploy a robust, fault-tolerant system that handles a two- to four-digit number of moderately complex tasks per second, with latencies between a few hundred milliseconds and 3~seconds. While high throughput requirements can only be handled with more expensive hardware, and costs for hardware increase super-linearly in maximum throughput required, the costs for moderate systems can be as low as $10^{-5}~$USD (provided considerable average utilization of capacity) or a small five-digit~USD figure annually even when running on on-demand cloud services. Nonetheless, \ac{hlf} reaches its limits for systems requiring a stable throughput of several thousand transactions per second or for systems with slightly lower throughput requirements and an additional need for a large two-digit number of nodes, privacy, or complex workloads. To meet such requirements, further improvements and specific features that allow optimizing the performance of \ac{hlf} nodes, such as the ones suggested by~\citet{thakkar2021scaling}, will be necessary.

These findings are of great relevance to supply chain management as the proposed blockchain solutions usually have high throughput, low latency, and global infrastructure requirements. Despite covering a comprehensive list of variables, however, the analytic focus of this study required it to leave certain areas unexplored. Future research may, for example, find it beneficial to compare the various supported programming languages, such as Go, Node, and Java. In doing so, we hope that future research will use our extended \ac{dlps} framework to examine additional implementations. Furthermore, while this paper considers a recent version of \ac{hlf} to evaluate current features, such as private data collections, it is important to note that the blockchain framework is still evolving comparatively fast, with developers focusing on further improving the overall performance along with new features like anonymous credentials and zero-knowledge proofs for improved transaction privacy. Therefore, although our spot-check comparisons between \ac{hlf} version 1.4 and
2.0 have indicated that the performance deviations in this update are only small, our analysis is only valid for particular release versions and requires additional testing once an update is introduced. Nonetheless, the list of impact parameters and the functionalities by which we extended the \ac{dlps} should be able to support analyses of future updates on the \ac{hlf} code. What is more, it supports other enterprise blockchains like Quorum, all of which offer similar functionalities (smart contracts and private transactions) and parameters (such as block time) in deployments with different network architectures and hardware.

Another promising avenue for future research is an examination of the impact of consensus mechanisms on updates of \ac{hlf}, and indeed on blockchain implementations in general. Our performance evaluation with RAFT consensus and the comparison with the Solo and Kafka ordering (that are deprecated in version 2.0 but available in 1.4) indicates that the ordering service is not currently a bottleneck in \ac{hlf}. However, this may change when a more fault-tolerant consensus mechanism is used, one that involves more communication, such as a three-phase commit in Practical Byzantine Fault Tolerant (PBFT). PBFT was previously used in \ac{hlf} version 0.6 but removed in version 1.0, making it impossible to extensively compare this consensus mechanism with others. However, there are ongoing efforts to provide PBFT consensus for \ac{hlf}, and since \ac{hlf} was designed to be sufficiently modular for various consensus mechanisms to be integrated, several should be available for analysis and comparison in the near future, be they assimilated from other Hyperledger projects (such as Proof of Elapsed Time in Hyperledger Sawtooth, Redundant BFT in Hyperledger Indy) or from other domains entirely (like HoneyBadger BFT or Solana’s Proof of History). To this end, it may prove helpful to use the support of \ac{dlps} to investigate latency sensitivity with benchmarking systems that are physically far distributed.

A further issue worth bearing in mind when embarking on future research is that, while researchers and practitioners focus on \ac{hlf}, we wish to reiterate the need to also investigate the potential of other blockchain implementations. Of course, the architecture of different blockchain systems differs in specific details, but the core architectural principles are common to most. A thorough consideration of architecture, setup, business logic, network, and robustness when benchmarking different blockchain implementations will ultimately facilitate a better comparison of results.

Already, the use and reach of blockchain technology in the industry have come a long way. Especially with the latest releases of \ac{hlf}, blockchain implementations have come much closer to operational maturity than ever before. In view of our own research, we are confident that pilot projects will soon be ready for operational implementation, ultimately improving supply chains. Nevertheless, our collective understanding of this technology, its performance, and scalability, in general, remains limited.  
As a research community, we have only just started to identify a comprehensive set of factors that impact this kind of distributed system. Gaining further insights will refine the design of blockchain infrastructures and applications, pushing the technological boundaries toward more advanced systems. To expedite this development, future research would do well to investigate alternative or complementary approaches to the provision of blockchains that facilitate high performance. This includes pursuing incremental performance optimizations, such as splitting certain workloads to dedicated nodes~\citep{thakkar2021scaling}, incorporating serverless implementations of nodes or even building completely serverless blockchains that can scale elastically on demand and reach far greater throughput at the tradeoff of increased centralization~\citep{sedlmeir2022serverless}, or reducing the computational and data complexity of the application itself. The latter could be achieved, for instance, by means of sharding or using succinct verifiable computation techniques, such as zero-knowledge proofs to reduce the workload for multiple nodes at the cost of increased (but heavily parallelizable) computation for one entity~\citep[see, e.g., ][]{simunic2021verifiable, ruckel2022fairness}. At present, succinctly verifiable computation is arguably more relevant for permissionless blockchains because the amortized complexity only pays off when there is a considerable number of validators that need to verify a transaction, but the alternative approaches suggested above may benefit large networks and a high number of endorsers. Our research has led us to believe that by combining these approaches, blockchains can ultimately provide even the scalability required by the most demanding enterprise applications.



\clearpage
\bibliographystyle{abbrvnat}
\bibliography{references.bib}



\end{document}